\documentclass[lettersize,journal]{IEEEtran}
\usepackage{amsmath,amsfonts}
\usepackage{algorithmic}
\usepackage{algorithm}
\usepackage{array}
\usepackage[caption=false,font=normalsize,labelfont=sf,textfont=sf]{subfig}
\usepackage{textcomp}
\usepackage{stfloats}
\usepackage{url}
\usepackage{hyperref}
\usepackage{verbatim}
\usepackage{graphicx}
\usepackage{color,soul}
\usepackage{cite}
\usepackage{authblk} 
\usepackage[table]{xcolor}
\def \figpath {./}

\begin{document}

\title{Prototyping and Test of the ``Canis'' HTS Planar Coil Array for Stellarator Field Shaping}

\author{D. Nash, D.A. Gates, W.S. Walsh, M. Slepchenkov, D. Guan, A.D. Cate, B. Chen, M. Dickerson, W. Harris, U. Khera, M. Korman, S. Srinivasan, C.P.S. Swanson, A. van Riel, R.H. Wu, A.S. Basurto, B. Berzin, E. Brown, C. Chen, T. Ikuss, W.B. Kalb, C. Khurana, B.D. Koehne, T.G. Kruger, S. Noronha, J. Olatunji, R. Powser, K. Tamhankar, K. Tang, A. Tarifa, M. Savastianov, J. Wasserman, and C. Yang}
\affil{Thea Energy, Inc., Kearny, NJ, USA}


\maketitle

\begin{abstract}
  Thea Energy, Inc. is currently developing the ``Eos'' planar coil stellarator, the Company’s first integrated fusion system capable of forming optimized stellarator magnetic fields without complex and costly modular coils. To demonstrate the field shaping capability required to enable Eos, Thea Energy designed, constructed, and tested the ``Canis'' 3x3 array of high-temperature superconductor (HTS) planar shaping coils after successfully demonstrating a single shaping coil prototype. Through the Canis 3x3 magnet array program, Thea Energy manufactured nine HTS shaping coils and developed the cryogenic test and measurement infrastructure necessary to validate the array's performance. Thea Energy operated the array at 20 K, generating several stellarator-relevant magnetic field shapes and demonstrating closed loop field control of the superconducting magnets to within 1\% of predicted field, a margin of error acceptable for operation of an integrated stellarator. The Canis magnet array test campaign provides a proof of concept for HTS planar shaping coils as a viable approach to confining stellarator plasmas.

\end{abstract}


\section{Introduction}
\IEEEPARstart{T}{hea Energy}, Inc. is developing a series of first-of-a-kind planar coil stellarators\cite{gates_stellarator_2025}, including a neutron source deuterium-deuterium (D-D) stellarator\cite{swanson_scoping_2025}, called Eos, and a deuterium-tritium (D-T) fusion pilot plant (FPP) stellarator that is a scale up of Eos, called Helios. The basis for this novel stellarator design has been exclusively licensed from a patent developed at the Princeton Plasma Physics Laboratory (PPPL)\cite{planar_coil_stellarator_patent}. The primary motivations behind this stellarator design approach are to minimize design and manufacturing complexity of the magnet coils and therefore system construction cost and timeline, and to enable a viable power plant maintenance scheme by making the shaping coils fully demountable.

Historically, stellarators have used what are frequently referred to as ``modular coils'', which are non-planar and often superconducting magnets which wrap around a plasma vessel of non-uniform curvature. Modular coils are complex and expensive to design and manufacture, for reasons including that they require tight geometric tolerances applied over large 3-dimensional shapes, and that they cannot be wound in tension due to their concavity. Building a 3-dimensional magnet coil system for a stellarator has been attempted by at least three public sector programs: Wendelstein 7-X at the Max Planck Institute for Plasma Physics\cite{bosch_lessons_2011,bosch_engineering_2018}, NCSX at Princeton Plasma Physics Laboratory\cite{neilson_lessons_2009,chrzanowski_lessons_2009}, and HSX at University of Wisconsin - Madison\cite{almagri_hsx_1999}. All three programs significantly exceeded cost and schedule estimates due in part to the complexity of the modular coils, with the entire NCSX program eventually being canceled.

The planar coil stellarator generates the required 3-dimensional fields by employing an array of planar encircling coils, providing axial magnetic field (similar to toroidal field coils in a tokamak), and an array of planar shaping coils which are tiled on the external surface of the vacuum vessel or other surrounding structure\cite{kruger_coil_2025}. The tile-like shaping coils can be grouped into field shaping units (FSUs) which form the basic maintainable and demountable unit of the field shaping system. Each FSU represents a modular assembly and can provide vacuum insulation, cooling, and power distribution for its constituent planar shaping coils. 

\begin{figure}[!h]
  \centering
  \includegraphics[width=3.2in]{\figpath/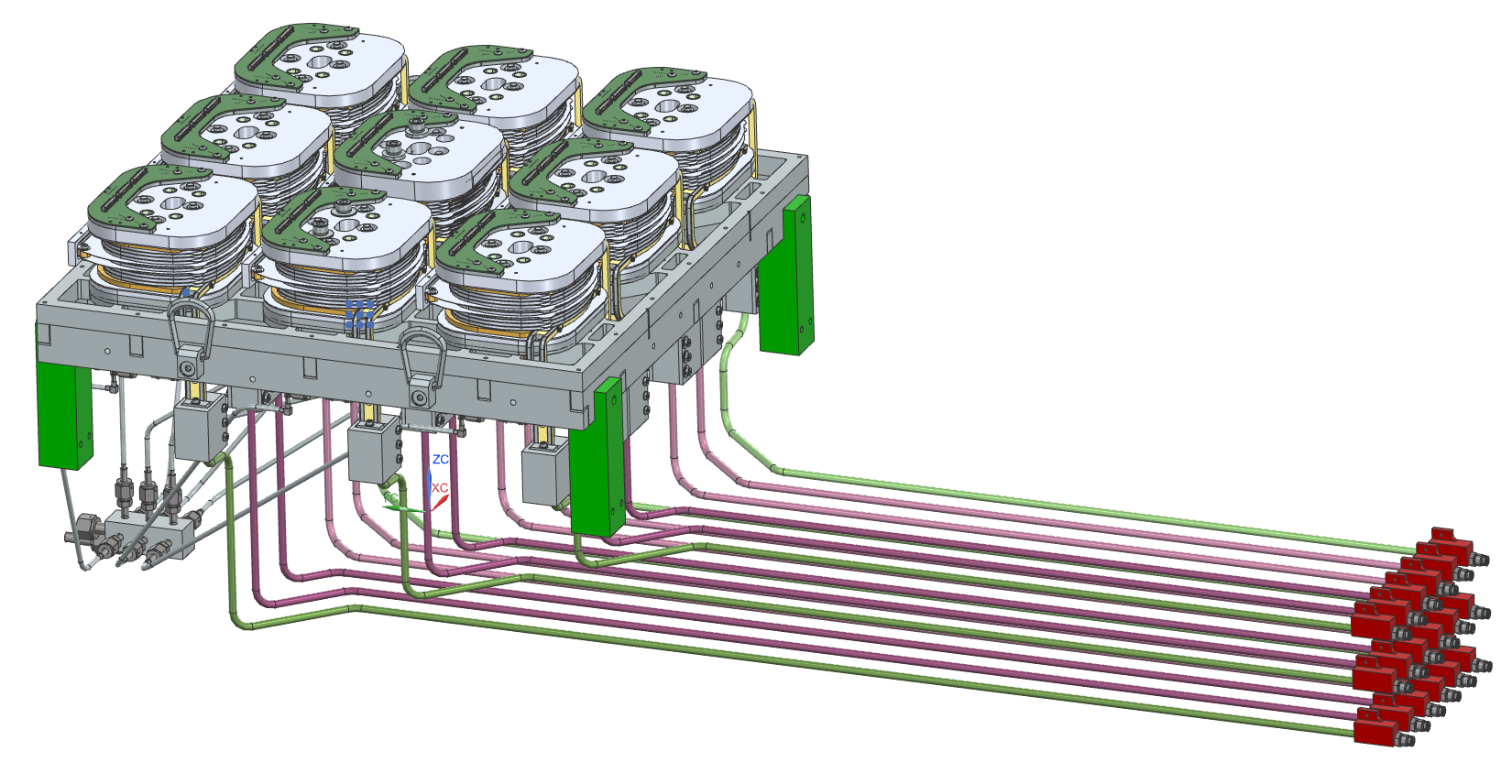}
  \caption{A CAD rendering of the Canis 3x3 magnet array, showing plumbing and current leads.}
  \label{fig_canis_render}
  \end{figure}

The encircling coils must generate a background field exceeding 10 T at some planar shaping coils to achieve the on-axis field strengths of 6 T at the plasma required for an optimized planar coil stellarator\cite{gates_stellarator_2025}. When also considering self-field from the shaping coils, field on the conductor in the shaping coils is expected to exceed 14 T. Rare-earth barium copper oxide (REBCO) was chosen as the superconductor to operate the shaping coils at these fields at near-steady state with a current density exceeding 200 A/mm$^2$ required by the application. REBCO superconductors have demonstrated the performance required for high-field, fusion-relevant magnet systems (20 T, 20 K) and are commercially available\cite{molodyk_development_2021}. REBCO coated conductors have now been demonstrated in several large-scale, fusion-relevant magnet and cable demonstrations, including the SPARC TFMC and CSMC magnet programs from Commonwealth Fusion Systems and MIT-PSFC\cite{hartwig_sparc_2023, sanabria_development_2024}.

Although REBCO magnet technology has matured steadily toward applications for fusion over the past decade, several critical criteria relevant to fusion applications remain. HTS coated conductors have a high specific energy, and a normal zone propagation velocity (NZPV) up to two orders of magnitude slower than low-temperature superconductors (LTS) \cite{schwartz_quench_2014}, in some cases making quench difficult to detect before damaging peak temperatures develop. REBCO coated conductors have also been shown to be susceptible to delamination in transverse stress and peeling modes \cite{laan_delamination_2007, lu_rebco_2025}, which can result from both electromagnet and thermal stresses. The significant anisotropy of in-field performance in REBCO conductors can also present a challenge to optimizing conductor usage and operating margin. Concerns related to controlling arrays of shaping coils largely center on the inductive and thermal coupling between adjacent magnets. Control systems must consider and compensate for mutual inductance between coils. 

Thea Energy undertook the design, manufacture, and test of a prototype shaping coil array, nicknamed ``Canis'', to derisk and demonstrate the fundamental field shaping capabilities of an array of HTS magnets. A rendering of the array is shown in Fig. \ref{fig_canis_render}. The primary objectives of the Canis magnet array are outlined in Table \ref{tab:objectives}. Program objectives related to \textit{Magnet Manufacturing Development} and \textit{Field Shaping Control Development} are discussed in this paper. Some remaining objectives in \textit{Eos Scenario Development} and \textit{Eos Shaping Coil and Field Shaping Unit (FSU) Design Criteria} may be discussed in future publications.

\newcolumntype{L}[1]{>{\raggedright\arraybackslash}p{#1}} 

\begin{table}[!h]
  \caption{Canis 3x3 magnet array program and test objectives\label{tab:objectives}}
  \centering
  \begin{tabular}{| L{1.8cm} | p{6.2cm} |}
  \hline
  \textbf{Group} & \textbf{Objectives}\\
  \hline
  Magnet Manufacturing Development &  \begin{itemize}
                              \item{Demonstrate \(\leq\)1 day double pancake (DP) takt time}
                              \item{Demonstrate \(\leq\)4 days total DP production time}
                              \item{Build array with HTS from \(\geq\)3 suppliers}
                              \end{itemize}\\
  \hline
  Field Shaping Control Development & \begin{itemize}
                              \item{Demonstrate closed-loop control of magnetic field at each magnet to generate $B_{z}$ field shapes}
                              \item{Reproduce several unique and Eos-relevant $B_{z}$ field shapes with \(\leq\)1\% RMS field error}
                              \item{Validate $B_{z}$ iso-surfaces by scanning magnetic field on a plane located 25 cm away from and parallel to the array midplane}
                              \end{itemize}\\
  \hline
  Eos Scenario Development & \begin{itemize}
                              \item{Demonstrate controlled transition between multiple field shapes, including Eos-relevant shapes}
                              \item{Demonstrate simultaneous discharge of the array in several high-stored energy configurations}
                              \item{Demonstrate non-propagation of coil quench in several quench configurations}
                              \item{Demonstrate insensitivity of control system to several classes of manufacturing defect}
                              \end{itemize}\\
  \hline
  Eos Shaping Coil and Field Shaping Unit (FSU) Design Criteria & \begin{itemize}
                              \item{Validate key electromagnetic and thermal magnet FEA models}
                              \item{Validate transient and steady-state system cooling predictions}
                              \item{Validate performance of hybrid current leads}
                              \end{itemize}\\
  \hline
  \end{tabular}
  \end{table}

\section{Planar Shaping Coil Design}
The application of the shaping coil magnet array toward a planar coil stellarator informed several key design criteria which would guide further development. Firstly, given the eventual integration of many hundreds of planar shaping coils and magnet current leads, the application encourages a low terminal current to enable dense and highly flexible current lead routing. A terminal current at which copper current leads and feedthroughs are feasible and commercially available was desired to enable non-superconducting options and to simplify design and integration. All-copper current leads have been demonstrated for low-current HTS magnet systems, including the 60 A and 120 A Large Hadron Collider (LHC) dipole corrector magnets operating at 1.9 K \cite{ballarino_1999}. The nominal operating current $I_{op}$ of the Canis shaping coil was selected to be 150 A, with maximum operating current $I_{max}$ of 250 A. Operating at 20 K, the minimum critical current of a Canis shaping coil was expected to exceed 600 A.

Secondly, the multi-year lifetime of a stellarator planar shaping coil and reliance on one or few parallel HTS conductors demanded a stable and defect-tolerant coil architecture. In 2G HTS tapes, latent manufacturing defects and handling induced degradation can result in a sharp, localized reduction in transport critical current, or ``$I_{c}$ dropouts''. Non-insulated (NI) and partially-insulated HTS magnets have been shown to be particularly robust to $I_{c}$ dropouts, by allowing current to bypass local series resistance through a turn-to-turn radial resistance, even at operating currents significantly exceeding the REBCO's critical current\cite{hahn_hts_2011}. Further, a rigid thermally and electrically conductive structural matrix was preferred over a dry winding in an effort to maintain a reliable radial resistivity over the lifetime of the coil. A dry winding process was not considered for this application due to the tendency for turn-to-turn contact resistance to change by up to an order of magnitude over many cycles\cite{lu_contact_2018}. The general robustness of the coil, particularly of the initial prototype, was of significantly higher importance than the overall current-to-field efficiency, engineering current density, or field homogeneity.

Third, the intended operational profile and test goals for the array required a coil design which could reliably sustain repeated instantaneous discharges, both planned and unplanned. In parallel, the dense packing of coils and current leads expected within the FSU encouraged a design without external quench protection hardware. From these two criteria, a passively safe, self-protecting coil architecture was desired. Non-insulated and partially-insulated HTS coils at low stored energy have been demonstrated to be passively safe and self-protecting during a quench \cite{hahn_no-insulation_2013, lecrevisse_metal-as-insulation_2018, song_over-current_2015, choi_characteristic_2016, seungyong_hahn_no-insulation_2012}, dissipating stored magnetic energy into the internal radial resistance path. The localized hotspot temperature following a quench can be minimized by appropriate selection of the turn-to-turn contact resitivity, $R_{ct}$\cite{lecrevisse_metal-as-insulation_2022}. For the Canis shaping coils, a soldered metal insulation architecture was chosen for its high thermal conductivity, lower thermal contraction risk (e.g. reduced layer delamination risk) due to its ductility, and target performance criteria available in literature\cite{mun_electrical_2020}.

The shaping coil was designed to use a double pancake (DP) architecture, wherein two single pancake (SP) layers were wound on a common bobbin and electrically connected by a resistive inner joint. DPs were then electrically connected using HTS outer joints. Coated metal ``interlayers'', designed to be electrically insulating but thermally conductive, were installed between all SP winding layers to insulate SPs while maximizing conductive heat transfer through the coil. Other design criteria related to the physical size and packing of the array were derived from a conceptual layout of the FSU, including the rounded rectangle coil shape, overall dimensions, and current lead geometry. A summary of key prototype shaping coil specifications are outlined in Table \ref{tab:table_coil_specs}.

\begin{table}[!t]
  \caption{Canis planar shaping coil specifications\label{tab:table_coil_specs}}
  \centering
  \rowcolors{2}{white}{gray!25}
  \begin{tabular}{p{4cm}|p{2.5cm}|p{1cm}}
  \hline
  \textbf{Specification} & \textbf{Value} & \textbf{Units}\\
  \hline
  Operating Current, nominal, $I_{op}$ & 150 & A\\
  \hline
  Operating temperature, nominal, $T_{op}$ & 20 & K\\
  \hline
  Number of double pancakes (DPs) & 5 & ---\\
  \hline
  Total turns, $N_{turns}$ & 1,500 & ---\\
  \hline
  Total current, nominal, $I_{tot}$ & 225 & kA\\
  \hline
  Stored energy, nominal, $E_{mag}$ & 4.4 & kJ\\
  \hline
  Winding pack (WP) current density, $J_{wp}$ & 180 & A/mm$^{2}$\\
  \hline
  Coil inductance, $L$ & 0.39 & H\\
  \hline
  Magnetic field strength in bore, $B_{z=0}$ & 1.8 & T\\
  \hline
  Peak field at $I_{op}$, $B_{self}$ & 3.7 & T\\
  \hline
  Coil shape & Rounded rectangle & ---\\
  \hline
  Winding dimensions L x W x H, nominal & 190 x 163 x 47 & mm\\
  \hline
  REBCO tape width & 4 & mm\\
  \hline
  Coil architecture & Soldered metal insulation (SMI) & ---\\
  \hline
  \end{tabular}
  \end{table}

Prior to manufacturing the nine shaping coils required for the Canis magnet array, an initial full single coil prototype was manufactured and tested\cite{tang_zethus_2025}. Test goals included characterization of coil electromagnetic parameters at 20 K, quantifying degradation over thermal and operational cycling, and characterizing the coil's quench response.

\section{Planar Shaping Coil Manufacturing}
The Canis magnet array required the manufacture, assembly, and test of nine complete winding packs (WPs), composed of 45 DPs. Thea Energy developed a manufacturing line capable of producing up to two DPs per day, with full WP assembly and liquid nitrogen acceptance testing taking one day each. Fabrication of the soldered metal insulation coils required the development of several key manufacturing processes and equipment, including the commissioning of a REBCO winding table and a solder impregnation process.

The winding machine developed for the production of planar shaping coils was designed to wind \(\geq\)2 tape spools---for HTS and other co-wound tapes---under a controlled tension ranging from 10 N to 60 N. In addition to tape tension, radial compaction was introduced during the winding process by a pneumatic roller-pusher, which applied radial pressure to the outer turns.

Insensitivity to HTS supplier was demonstrated by composing the Canis magnet array from WPs wound from three different REBCO suppliers, including manufacturers of yttrium barium copper oxide (YBCO) and gadolinium barium copper oxide (GdBCO). HTS wire from Supplier 1 and Supplier 2 had a 5 $\mu$m copper plating layer on either side of the tape, while Supplier 3 had a 20 $\mu$m layer.   

Solder impregnation was accomplished by co-winding a solid, low-temperature solder alloy in with HTS and other layers during the winding process using a novel technique. Following winding, the full DP was reflowed at 150 \textdegree{}C to melt the low-temperature solder alloy and activate flux which had also been introduced during the winding process. After reflow, the coils were cleaned by immersion into an ultrasonic alcohol bath. The baseline manufacturing process was not altered or tailored for any one supplier, resulting in variation in electromagnetic properties, particularly of radial resistance.

\begin{figure}[!h]
  \centering
  \includegraphics[width=3.5in]{\figpath/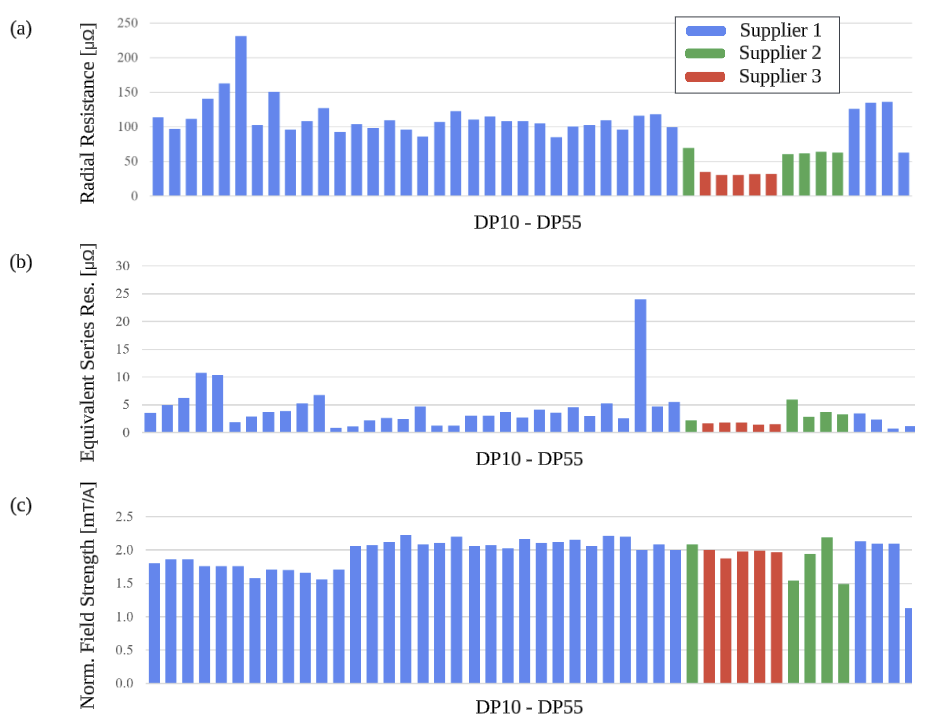}
  \caption{Double pancake (DP) performance trends over serial number as measured in liquid nitrogen testing at 77 K, for DPs manufactured from three different suppliers. Shown are (a) radial resistance, (b) equivalent series resistance (ESR) and (c) field strength normalized by power supply current. Field is measured by a Lake Shore HGCA-3020 cryogenic Hall effect sensor, located along the central axis of the coil 36.3 mm from the coil midplane.}
  \label{fig_dp_trends}
  \end{figure}

All DPs were electrical tested in liquid nitrogen using either a stepped or ramped current profile to \(\geq\)80 A. The minimum $I_{c}$ expected in any DP at 77 K was predicted to be \(\geq\)45 A, and each coil was charged and characterized beyond the first local normal zone transitions in the winding. Each DP was characterized for equivalent series resistance (ESR), radial resistance, and magnetic field strength normalized by power supply current. Performance trends of the DPs over serial number is shown in Fig. \ref{fig_dp_trends}.

\begin{figure}[!h]
  \centering
  \includegraphics[width=3.5in]{\figpath/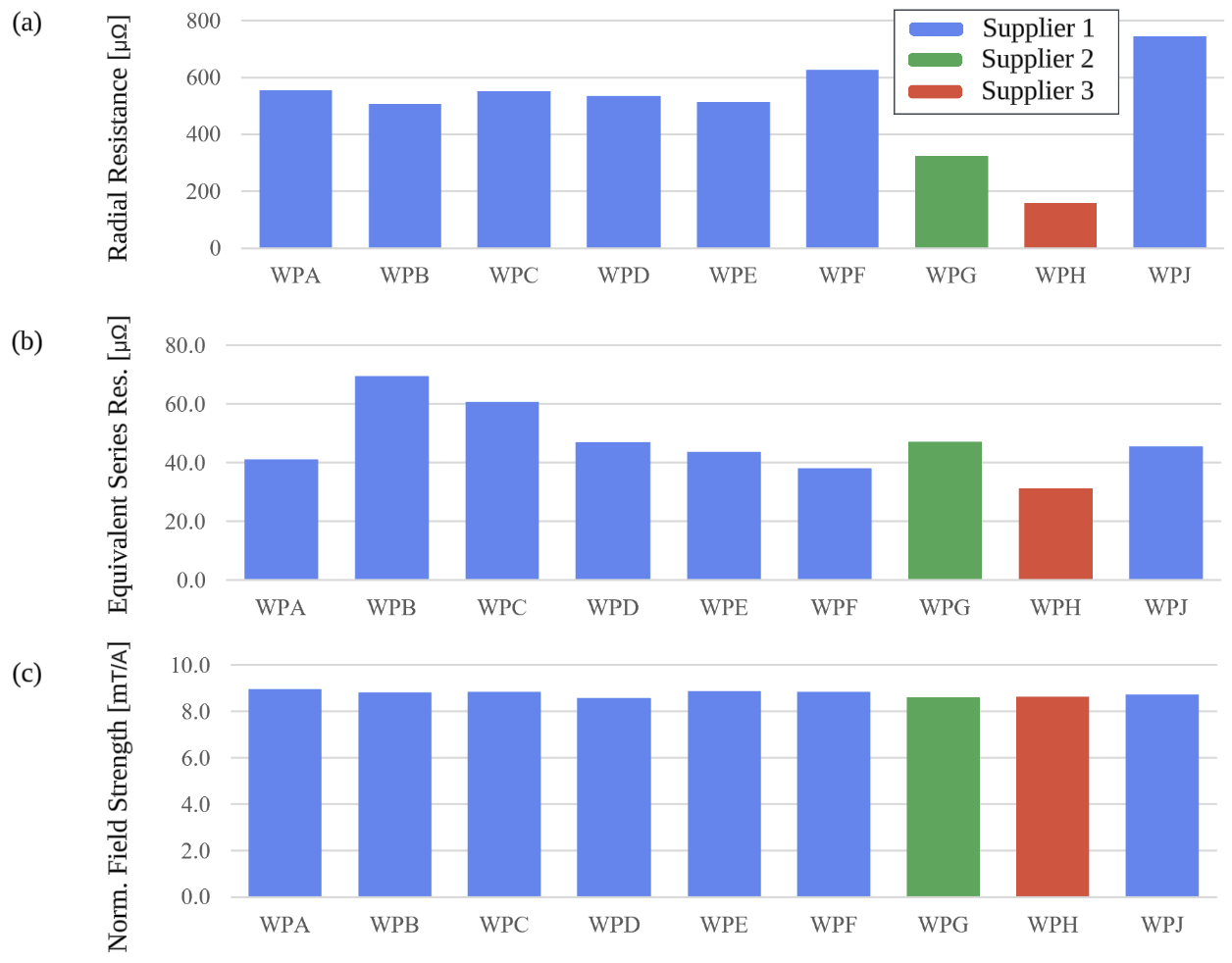}
  \caption{Winding pack performance trends over serial number as measured in liquid nitrogen testing at 77 K. Shown are (a) radial resistance, (b) equivalent series resistance (ESR) and (c) field strength normalized by power supply current. Field is measured by a Lake Shore HGCA-3020 cryogenic Hall effect sensor, located along the central axis of the coil 36.3 mm from the coil midplane.}
  \label{fig_wp_trends}
  \end{figure}

DPs were grouped into nine WPs (serialized A-J, e.g. ``WPA'') by attempting to match radial resistance within 15\%, though WPD and WPJ did not meet this criteria. DPs were then arranged in stacking order by a manual process of matching the outer radial dimension of each wind winding (to minimize shimming), and attempting to locate higher ESR (and therefore Joule heating) windings in the center of WP where the highest $I_{c}$ were expected. WPG and WPH were composed only of DPs wound from HTS from Supplier 2 and Supplier 3, respectively.

WPs were then re-tested as complete assemblies in liquid nitrogen at 77 K. Each WP underwent the same ramp profile to 80 A from the DP tests, though for the complete WP, minimum $I_{c}$ was expected to be \(\geq\)36A. Performance trends of the WPs over serial number is shown in Fig. \ref{fig_wp_trends}.

\section{Canis 3x3 Magnet Array Design}

\begin{figure}[!h]
  \centering
  \includegraphics[width=3.2in]{\figpath/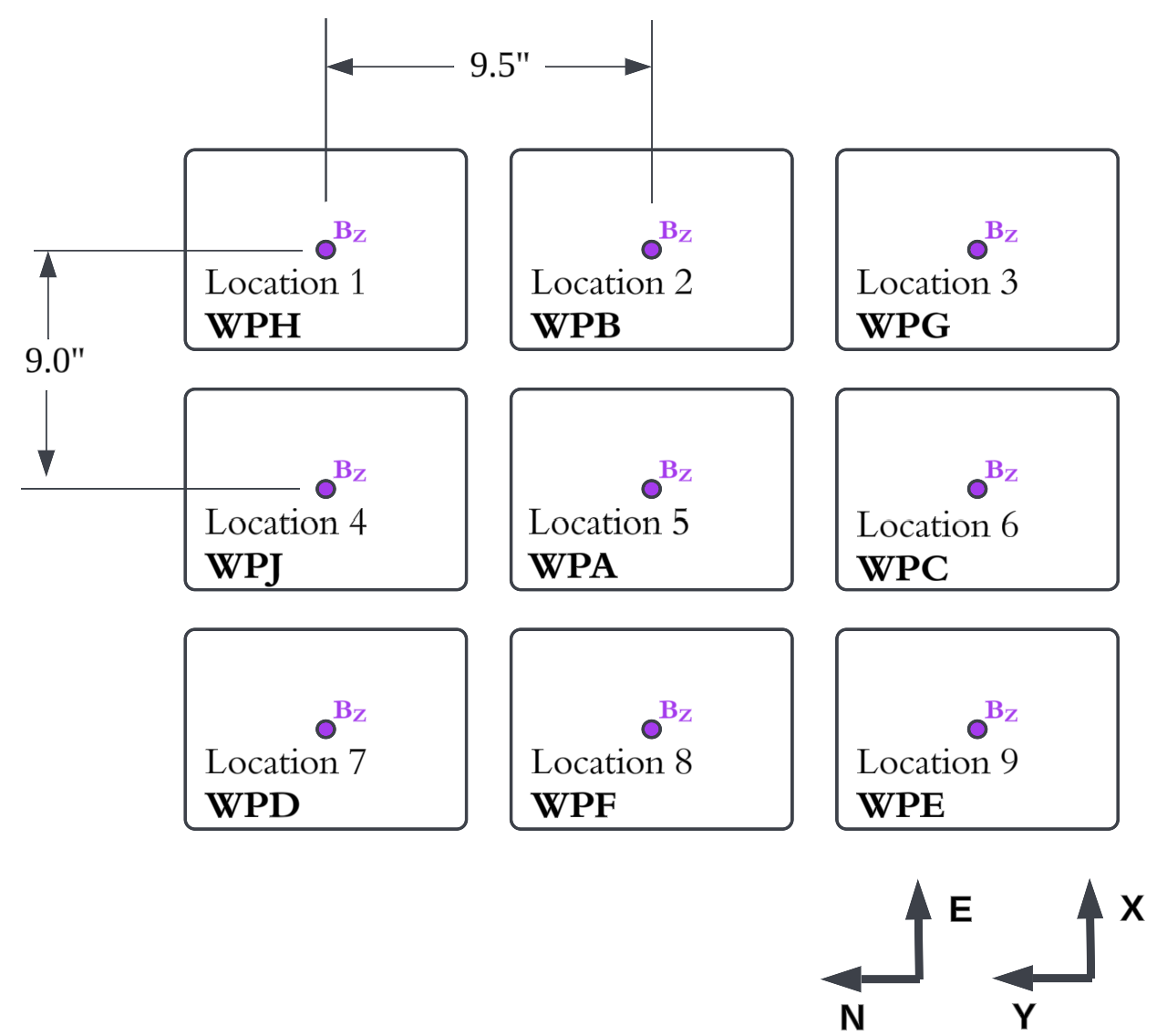}
  \caption{Canis 3x3 magnet array layout, showing mounting pitch, location indexing, and WP allocation. Also shown is the location of each Hall effect sensor.}
  \label{fig_array_layout}
  \end{figure}

The Canis magnet array is composed of nine prototype planar shaping coils as described above, tightly packed into a rectangular grid, with a pitch spacing in the X and Y directions of 9.0" and 9.5", respectively. The Canis magnet array layout is summarized in Fig. \ref{fig_array_layout}. Key performance specifications are outlined in Table \ref{tab:table_array_specs}.

\begin{table}[!t]
  \caption{Canis 3x3 magnet array specifications\label{tab:table_array_specs}}
  \centering
  \rowcolors{2}{white}{gray!25}
  \begin{tabular}{p{4cm}|p{2.5cm}|p{1cm}}
  \hline
  \textbf{Specification} & \textbf{Value} & \textbf{Units}\\
  \hline
  Operating current, nominal, $I_{op}$ & 150 & A\\
  \hline
  Maximum current, $I_{max}$ & \(\pm\)250 & A\\
  \hline
  Operating temperature, nominal, $T_{op}$ & 20 & K\\
  \hline
  Array arrangement & 9 magnets in 3x3 grid & ---\\
  \hline
  Magnet mounting pitch & 9.0 x 9.5 & inches\\
  \hline
  Stored energy at $I_{op}$, $E_{array}$ & 34.5 & kJ\\
  \hline
  Magnetic field at $I_{op}$ on plane 25 cm away, $B_{z=25cm}$ & 87.1 & mT\\
  \hline
  Maximum field on HTS at $I_{op}$, $B_{max,150A}$ & 3.04 & T\\
  \hline
  Maximum achievable field on HTS at $I_{max}$, charged in ``bullseye'' pattern, $B_{max,250A}$ & 5.72 & T\\
  \hline
  \end{tabular}
  \end{table}

The shaping coils in the array are coupled electromagnetically through mutual inductance, coupled structurally through the array plate, and coupled thermally via conduction through structure and a shared helium cooling loop. The mutual inductance matrix $L_{m,n}$ was derived via FEA simulations using COMSOL Multiphysics, and is shown below in Eq. (\ref{mat_L}) (row and column indexes refer to location indexes in Fig. \ref{fig_array_layout}, units are in [mH]):

\[
L_{m,n} = 
\begin{bmatrix}
L_{1,1} & L_{1,2} & \cdots & L_{1,8} & L_{1,9} \\
L_{2,1} & L_{2,2} & \cdots & L_{2,8} & L_{2,9} \\
\vdots & \vdots & \ddots & \vdots & \vdots \\
L_{8,1} & L_{8,2} & \cdots & L_{8,8} & L_{8,9} \\
L_{9,1} & L_{9,2} & \cdots & L_{9,8} & L_{9,9} \\
\end{bmatrix} =
\]

\begin{equation}
\scalebox{0.65}{  
$\begin{bmatrix}
370.37	& -9.83	& -1.04	& -9.59	&	-2.95	&	-0.72	&	-0.91	&	-0.66	&	-0.33 \\
-9.83	&	370.37	&	-9.83	&	-2.95	&	-9.59	&	-2.95	&	-0.66	&	-0.91	&	-0.66 \\
-1.04	&	-9.83	&	370.37	&	-0.72	&	-2.95	&	-9.59	&	-0.33	&	-0.66	&	-0.91 \\
-9.59	&	-2.95	&	-0.72	&	370.37	&	-9.83	&	-1.04	&	-9.59	&	-2.95	&	-0.72 \\
-2.95	&	-9.59	&	-2.95	&	-9.83	&	370.37	&	-9.83	&	-2.95	&	-9.59	&	-2.95 \\
-0.72	&	-2.95	&	-9.59	&	-1.04	&	-9.83	&	370.37	&	-0.72	&	-2.95	&	-9.59 \\
-0.91	&	-0.66	&	-0.33	&	-9.59	&	-2.95	&	-0.72	&	370.37	&	-9.83	&	-1.04 \\
-0.66	&	-0.91	&	-0.66	&	-2.96	&	-9.59	&	-2.95	&	-9.83	&	370.37	&	-9.83 \\
-0.33	&	-0.66	&	-0.91	&	-0.72	&	-2.95	&	-9.59	&	-1.04	&	-9.83	&	370.37 \\
\end{bmatrix}$
}
\label{mat_L}
\end{equation}

Structurally, all nine magnets were mounted to a single stainless steel structural array plate, described in Section 5B. All intercoil electromagnetic forces were fully reacted within the structural array plate. On the array plate, each magnet is mounted independently and cooled from parallel helium circuits. Each magnet is also powered by an independent power supply with its own pair of current leads as described in Section 5D. All magnets in the array were equipped with a 300W quench heater, installed on specially designed interlayers that evenly distributed heat into the central DP.

\section{Test Facilities}
\subsection{Cryostat}
The cryostat for the Canis magnet array and the supporting systems was sized to be 50\% larger than required for the experiment to be forward compatible with future magnet array systems. The diameter of the required vessel was calculated to be 80” with a height requirement of 20”. The cryostat vacuum vessel was recommissioned from an aluminum vessel which had been previously used in cosmic-ray detection experiments at Mt. Evans in the 1960s\cite{erickson_multiparticle_1970} to minimize lead time and cost. The height was extended 25” by welding eight arc sections to the pre-existing flanges from the old vessel to meet the requirements of current and future use. Each arc-section also contained an ISO-400 feedthrough port for subsystem feedthroughs and interconnects. In total, the cryostat provided eight ISO-400 flanges for vacuum, power, cooling, and instrumentation feedthroughs.

Structural simulations of vacuum loading indicated a factor of safety (F.O.S.) of \(\geq\)7, and the recommissioned vessel was qualified through several cycles of sustained vacuum loading with target base pressure \(\leq\)1x10$^{-6}$ torr at both the manufacturing facility and experimental facility. During the experiment, a cryostat vacuum pressure of \(\leq\)1x10$^{-5}$ torr was achieved and maintained using two turbomolecular pumps, following several purge cycles with nitrogen gas to remove moisture from internal surfaces and reduce overall pump-down time. 

The Canis cryostat employed a Copper-100 sheet metal and multi-layer insulation (MLI) thermal shield which enveloped the magnet array and support systems to mitigate radiative heat transfer from the walls of the vessel. FEA of the thermal shield indicated a reduction of radiated heat load to the 20 K cold mass from almost 5000 W without the shield to 22.5 W with the shield. Cooling of the thermal shield was achieved by a network of liquid nitrogen cooling lines thermally anchored to support structure and shield panels at various locations with a pressure driven, variable flow rate to allow for dynamic cooling control.

All aluminum and copper cryostat systems were analyzed for quench-induced deformation using ANSYS simulations with a target F.O.S for all bodies greater than 2. Peak deformation and stresses were determined by modeling the transient Lorentz body forces due to eddy currents on conductive structures during magnet array discharges with a quench-relevant time constant.

\subsection{In-Vessel Structures}
The structure directly supporting the Canis magnet array consisted of a SS316 plate with lateral stiffener bars, which was thermally stood off from a warmer structure by four G-10 vertical gravity supports. Stainless steel was selected for its stiffness and thermal resistance. The structural plate was designed to tightly control deformation from gravity loads and from steady-state and transient magnetic loads between WPs. Displacements due to gravity and Lorentz loads within the complete magnet array were limited to \(\pm\)0.2 mm to ensure that magnet position and magnetic field distribution were acceptably repeatable and invariant to electromagnetic deflections. A repeating cutout pattern was added to the structural plate to increase thermal resistance between WPs, reduce mass, and provide clearance for cooling and current lead connections. Array deflection from intermagnetic and gravity loads, and the transient thermal response of the array were estimated by FEA multi-physics simulation.

Gravity supporting structure was designed to minimize deformation of the array plate during testing and to mitigate conductive heat leak to the vacuum vessel. Two parallel, horizontal W10X12 SS316 beams supports were used to allow for fixturing to the vertical cylinder of the cryostat, with the volume between them used for current and cooling routing underneath each WP.

\begin{figure}[!h]
  \centering
  \includegraphics[width=3.2in]{\figpath/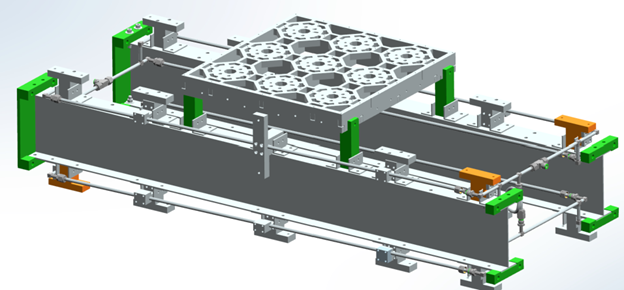}
  \caption{CAD rendering of in-vessel structure, including magnet array structural plate and gravity supports.}
  \label{fig_canis_structure}
  \end{figure}

Each G-10 vertical gravity supports was fixtured with a clevis to provide minimally constrained support of the magnet array structural plate during thermal contraction. The end of each parallel beam was supported by a G-10 cradle, with one end being bolted and the other a sliding contact, to mitigate thermal contraction of the beam supports and to minimize heat leak from the vacuum vessel. The horizontal gravity supports are actively cooled by liquid nitrogen. A CAD rendering of the in-vessel structures is shown in Fig. \ref{fig_canis_structure}. 

\begin{figure}[!h]
  \centering
  \includegraphics[width=3.2in]{\figpath/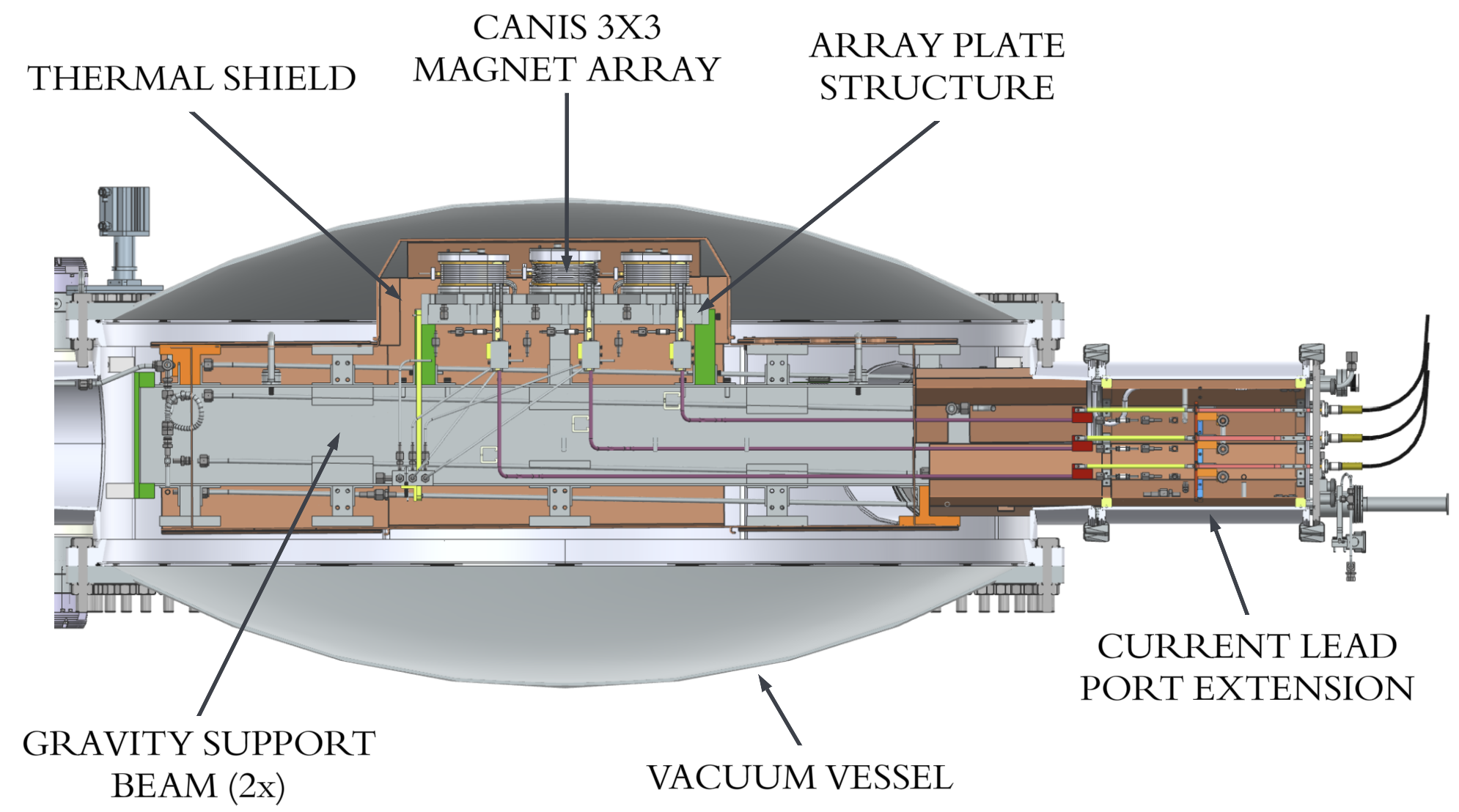}
  \caption{CAD rendering of cryostat section view, showing major in-vessel systems.}
  \label{fig_cold_mass}
  \end{figure}

\subsection{Cryogenic Cooling Systems}
Cryogenic cooling of the Canis magnet array and in-vessel support systems was provided by an 80 K liquid nitrogen (LN2) cooled 1st stage and a 20 K 2nd stage cooled by supercritical helium. The 2nd stage helium loop operated at a nominal pressure and temperature of 20 bar and 20 K, respectively. Supercritical helium was the coolant of choice due to its phase stability and fluid uniformity near the operating conditions, lower frictional pressure losses than liquid phase, and higher cooling power than gas phase. Helium is cooled and circulated by an ex-vessel cryoplant supplied by Absolut Systems with 4 Gifford-McMahon cold heads and a cryogenic circulation fan to drive flow, which was plumbed to conductively cool magnets, support structure, and HTS leads. Each magnet was mounted to a ``spiral'' cooling plate machined with internal helium cooling channels to provide uniform conductive cooling to a single side of the magnet. Flow rate was passively regulated to 20 g/s by an orifice located at the cryostat inlet. The helium cryoplant was tested at Thea Energy for heat rejection capacity and demonstrated 325 W of cooling power at 20 bar and 20 K with 60 Hz AC supply voltage. This tested capacity exceeded the predicted cooling power requirement of 135 W from steady-state and transient magnet heat loads, and from radiative and conductive heat leak to the 2nd stage.

The cryoplant was instrumented by the supplier to measure temperatures at the cryoplant inlet, outlet, and each cold head heat exchanger. The cryoplant was also instrumented to measure absolute pressure of the helium circuit, and differential pressure across the fan by which flow rate could be estimated. In-vessel instrumentation of the 2nd stage within the Canis cryostat consisted of Cernox\textsuperscript{\textregistered} RTDs on the cryostat inlet, outlet, and 20 K structure, pressure transducers on the inlet and outlet, and a cryogenic turbine flow meter on the cryostat outlet for direct flow rate measurement. 

For the 80 K 2nd stage, LN2 was supplied at 22 psig and plumbed to a phase separator at 1 atmosphere which provided gravity-driven flow through the in-vessel cryostat plumbing. Within the cryostat, the LN2 tubes were mounted to structures, thermal shield panels, and the warm end of the HTS leads to provide conductive cooling. Flow exits the cryostat and returns to the phase separator for venting or recirculation. Liquid nitrogen was the chosen for 1st stage cooling due to its low cost and robust supply chain that comes ready for use as is. LN2 cooling is particularly effective during cooldown due to its latent heat of vaporization, which increases cooling power until adequately low temperatures are reached. 

The flow of LN2 through in-vessel plumbing was controlled by a cryogenic proportional control valve (PCV) upstream of the cryostat, and a tube heater installed near the LN2 outlet that increased boil-off at the warm end to increase flow at the cryostat inlet. At steady-state conditions, the nominal system flow rate was 23.8 lb/hr with an estimated maximum cooling capacity of 17.8 kilowatts. The phase separator elevation induced $\sim$4 psig of head pressure upstream of the cryostat inlet, resulting in an inlet LN2 temperature of $\sim$80 K. The LN2 system was instrumented for inlet pressure and RTDs to measure temperature at the inlet, pre-LN2 heater outlet, and cryostat outlet.

\begin{figure*}[!h]
  \centering
  \includegraphics[width=\textwidth]{\figpath/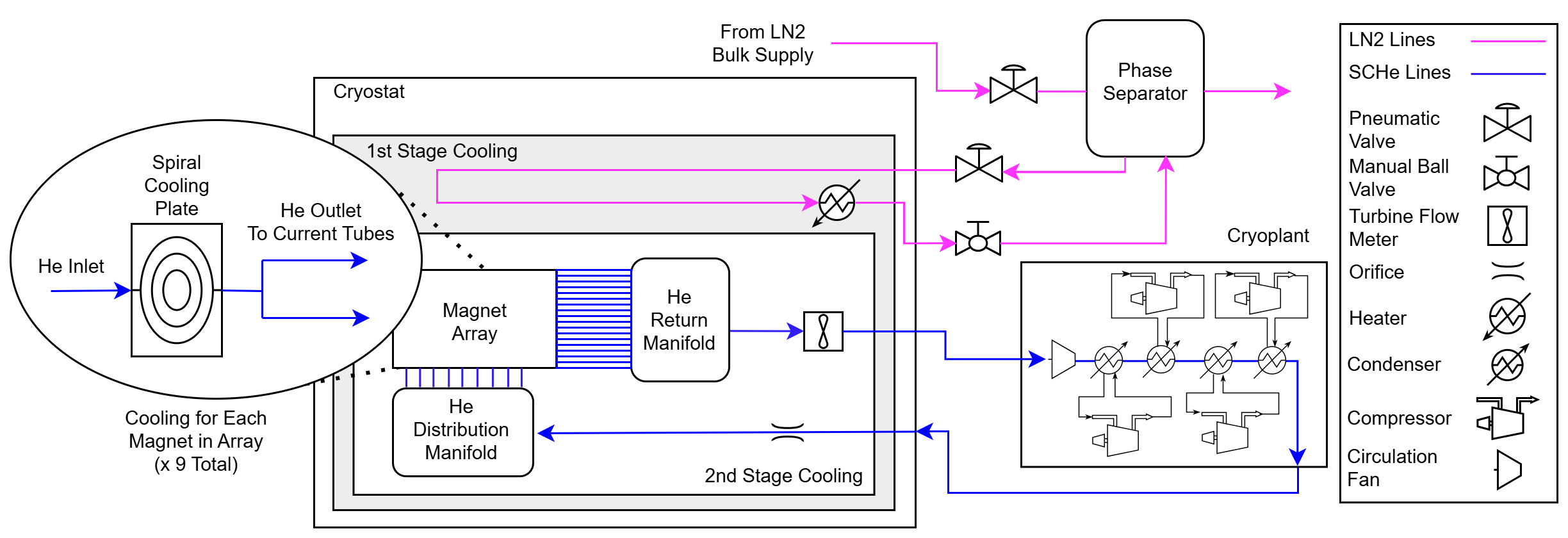}
  \caption{Simplified schematic of the Canis cryogenic fluid network, showing major components of the liquid nitrogen (LN2) and supercritical helium (ScHe) cooling loops.}
  \label{fig_fluids_network}
  \end{figure*}

\subsection{In-vessel Current Leads}
\begin{figure}[!h]
  \centering
  \includegraphics[width=3.5in]{\figpath/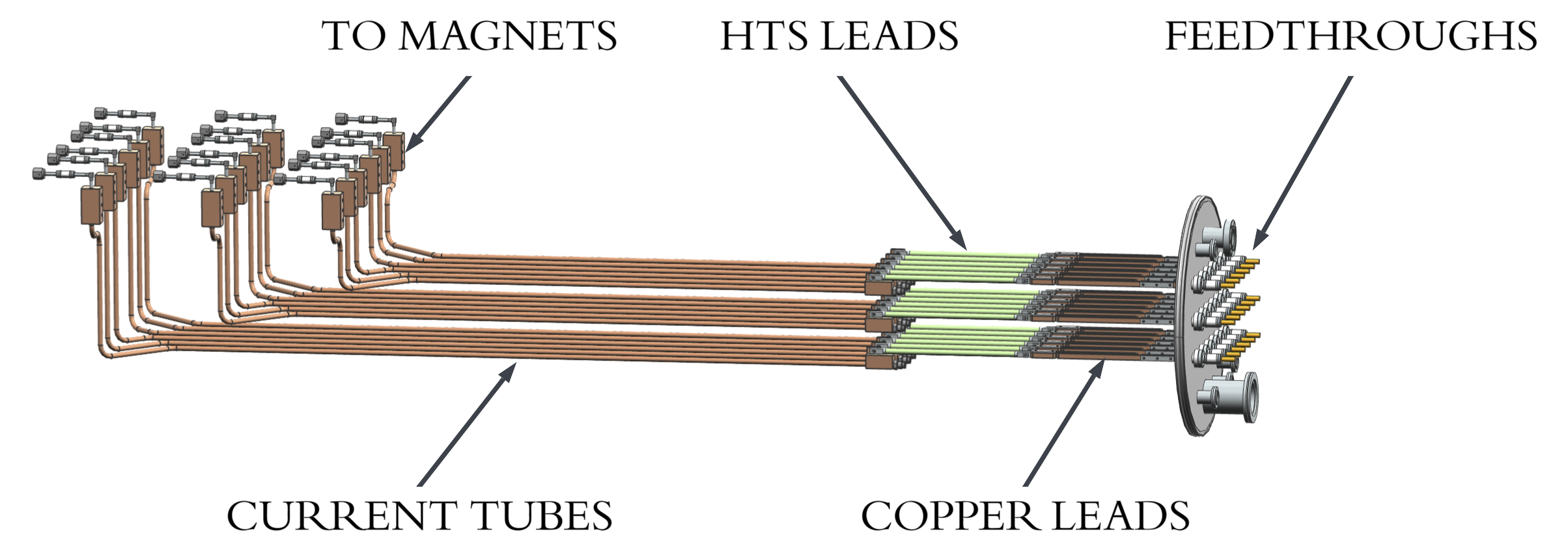}
  \caption{CAD rendering of hybrid HTS-copper in-vessel current leads.}
  \label{fig_current_leads}
  \end{figure}

Eighteen hybrid copper-HTS current leads were designed and manufactured in-house to route power from the feedthroughs to each Canis shaping coil. As shown in Fig. \ref{fig_current_leads}, the current leads consist of two resistive copper segments on either side of a COTS HTS superconducting lead with a 250 A current rating at 64 K. The copper leads routed to feedthroughs on the warm side of the HTS leads are actively cooled by conduction to a LN2-cooled copper block. On the 20 K side of the HTS leads, the current leads are composed of 3/8" C101 copper tubes (called ``current tubes'') cooled by flow of supercritical helium from the magnets. The relatively low operating current of the magnets (150 A) allows for the use of cryogenic copper tubes as current leads, which can be simply bent and brazed, greatly simplifying integration challenges associated with current lead routing. The HTS lead is conductively cooled by the copper leads on either end. This configuration results in temperatures of 80 K on the warm end of the superconducting lead, requiring 200 W of heat rejection by LN2 circulation. In total, the copper current tubes on the 20 K side of the HTS leads generated a combined \(\leq\)5 W of Joule heating at $I_{op}$, and a combined \(\leq\)1 W of heat leak was conducted through the HTS leads to the 20 K stage.

Voltage was measured across each HTS lead and current tube, and temperature measurements were captured on the cold and warm ends of the positive HTS leads. The current tubes were designed to withstand expected worst-case displacements due to Lorentz loads. The maximum von Mises stress calculated on the copper segment nearest to the magnet array was less than 9 MPa at 20 K with 0.5 mm displacement on the x-axis.

\subsection{Magnet and In-Vessel Instrumentation}
Individual WPs in the Canis magnet array were instrumented with voltage, temperature, and magnetic field measurements. For a modular design, each WP was assembled and acceptance tested with the same set of voltage taps, which included all pancake layers and resistive segments. For 20 K testing however, two different measurement resolution schemes were implemented to allow for finer characterization of some WPs while minimizing overall feedthrough requirements. A printed circuit board (PCB) mounted atop the WP allowed for selection of measurement resolution and a connectorized WP instrumentation interface. Two or four Cernox\textsuperscript{\textregistered} RTDs and a Lake Shore\textsuperscript{\textregistered} HGCA-3020 Hall effect sensor were installed on all WPs.

All instrumentation signals from WPs and other in-vessel systems were collected and routed to feedthroughs by a single in-vessel Canis Instrumentation Board (CIB). Other in-vessel signals included four-wire PT100 RTDs on the support structure and LN2 systems, Cernox\textsuperscript{\textregistered} RTDs placed on the magnet array structural plate and other 20 K systems, voltages and temperatures of the current leads, and a cryogenic turbine flow meter on the super-critical helium cryoplant return line. Over 800 signals were routed from the CIB to DB50 connectors on the cryostat instrumentation flanges. A summary of instrumentation used in the test campaign is shown in Table \ref{tab:sensors}.

\begin{table}[h]
  \caption{\label{tab:sensors} Canis 3x3 magnet array instrumentation summary}
  \centering
  \begin{tabular}{l|c|c|c}\hline
  Sensor Type & \multicolumn{3}{c}{Number of Channels}\\
    & High Res WP & Low Res WP & Support Systems \\\hline
  Voltage Tap & 18 & 13 & 27\\
  Cernox\textsuperscript{\textregistered} RTD & 4 & 2 & 17\\
  PT100 RTD & 0 & 0 & 23\\
  Hall Effect Sensor & 1 & 1 & 0\\
  ScHe Flow Sensor & 0 & 0 & 1 
  \end{tabular}
  \end{table}

\subsection{Power Supply System}
The nine magnets in the Canis magnet array were powered by nine independent Magna Power SL10-250 DC power supplies, capable of a maximum voltage and current output of 10 V and 250 A, respectively. A power conditioning circuit in an external unit in parallel with the power supply output was introduced to reduce output ripple and high frequency noise. This unit, called the ``coupling box'', also includes a voltage divider ground reference for the output, a high-accuracy current feedback sensor, and mechanical relays in an “H-Bridge” configuration that allow for commutation of the output current polarity to each magnet. The relatively low operating current of the Canis HTS magnets enables the use of inexpensive COTS relays for current commutation. The relays in the coupling box also provide isolation between each magnet and its power supply in the event of a quench or other fault condition. 

\subsection{Control Systems}
The Canis Control \& Data Acquisition System was an integrated system designed to operate and monitor the test cell, safety systems, and experiment via custom HMIs. It was composed of:

\begin{itemize}
  \item{A high-criticality HMI for testing operations}
  \item{A LabVIEW based control system executing real-time control algorithms to operate and monitor peripheral devices, either point-to-point or over a network}
  \item{A Field Shape Control System (FSCS), built on a Speedgoat real-time controller, which calculates reference currents for the Power Supply Units based on magnetic field sensor feedback}
  \item{A PLC-based safety system that continuously monitors environmental and connectivity signals to ensure safe operation}
  \item{A Linux-based telemetry logging server that captures all relevant measurements from the system}
  \end{itemize}

The primary National Instruments (NI) cRIO controller ran a Real-Time Linux OS with a user-programmable FPGA providing deterministic performance, modular I/O, and robust low-level control. The NI hardware acquired data from instrumentation outlined in Section 5E. It also gathered data from peripheral equipment and field I/O for real-time display, logging, and analysis. 

The magnetic field shape control algorithm in FSCS governed nine independent Power Supply Units driving the nine shaping coil magnets, maintaining a precise reference field shape.

\subsection{ATLAS Field Scanning Diagnostic}
A 2-dimensional gantry mounted with a high-accuracy 3-axis hall probe (nicknamed ``ATLAS'') was used to scan the resulting B-field at 25 cm above the midplane of the magnet array. The precise alignment of an ATLAS scan with respect to the magnet array required calibration and is discussed in section 6A. The ATLAS field scanning system measured magnetic field vector over a square area with 1.6 m sides with a grid spacing of 12.5 mm. Field measurements with accuracy of \(\pm\)0.1\% of full-scale reading were required. The Metrolab THM1176-MF Hall effect probe was selected as the sensor of choice, and came from the supplier pre-calibrated to \(\pm\)0.1\% accuracy along each axis up to 0.1 T.\phantomsection\label{hall_cal_footnote}\footnote{0.1\% calibration was a non-standard option from the supplier and was specially requested for this application.}

\begin{figure}[!h]
  \centering
  \includegraphics[width=3.2in]{\figpath/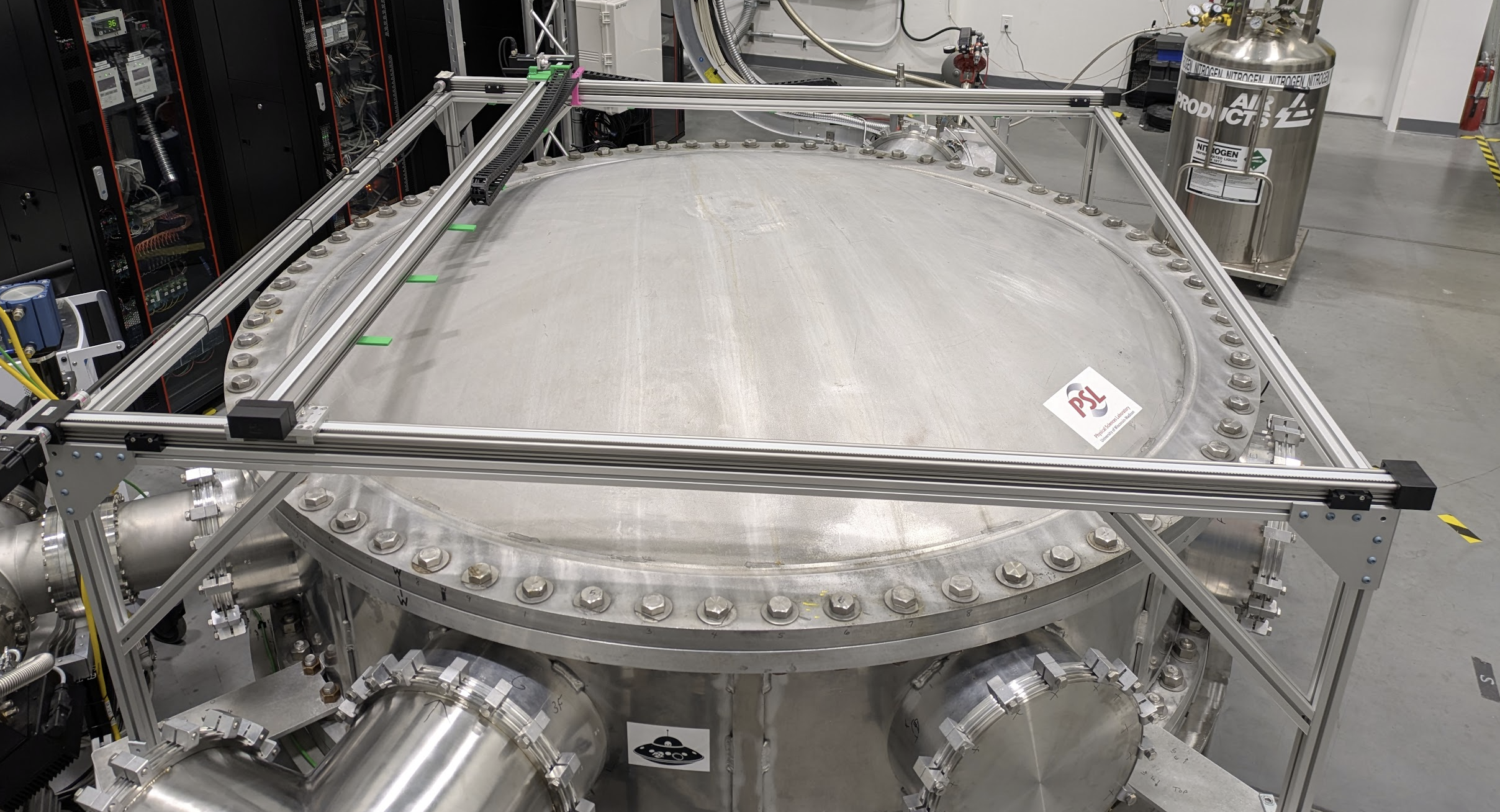}
  \caption{Photo of the ATLAS field scanning gantry system mounted on the Canis cryostat.}
  \label{fig_atlas}
  \end{figure}

Each axis of the ATLAS gantry was equipped with a high accuracy linear encoder (RKLC40-S) capable of \(\pm\)15 $\mu$m accuracy for precise localization of the scanning head. A photo of the ATLAS field scanner mounted on the Canis cryostat is shown in Fig. \ref{fig_atlas}.

\section{Shaping Coil Array Testing}

\begin{figure}[!h]
  \centering
  \includegraphics[width=3.2in]{\figpath/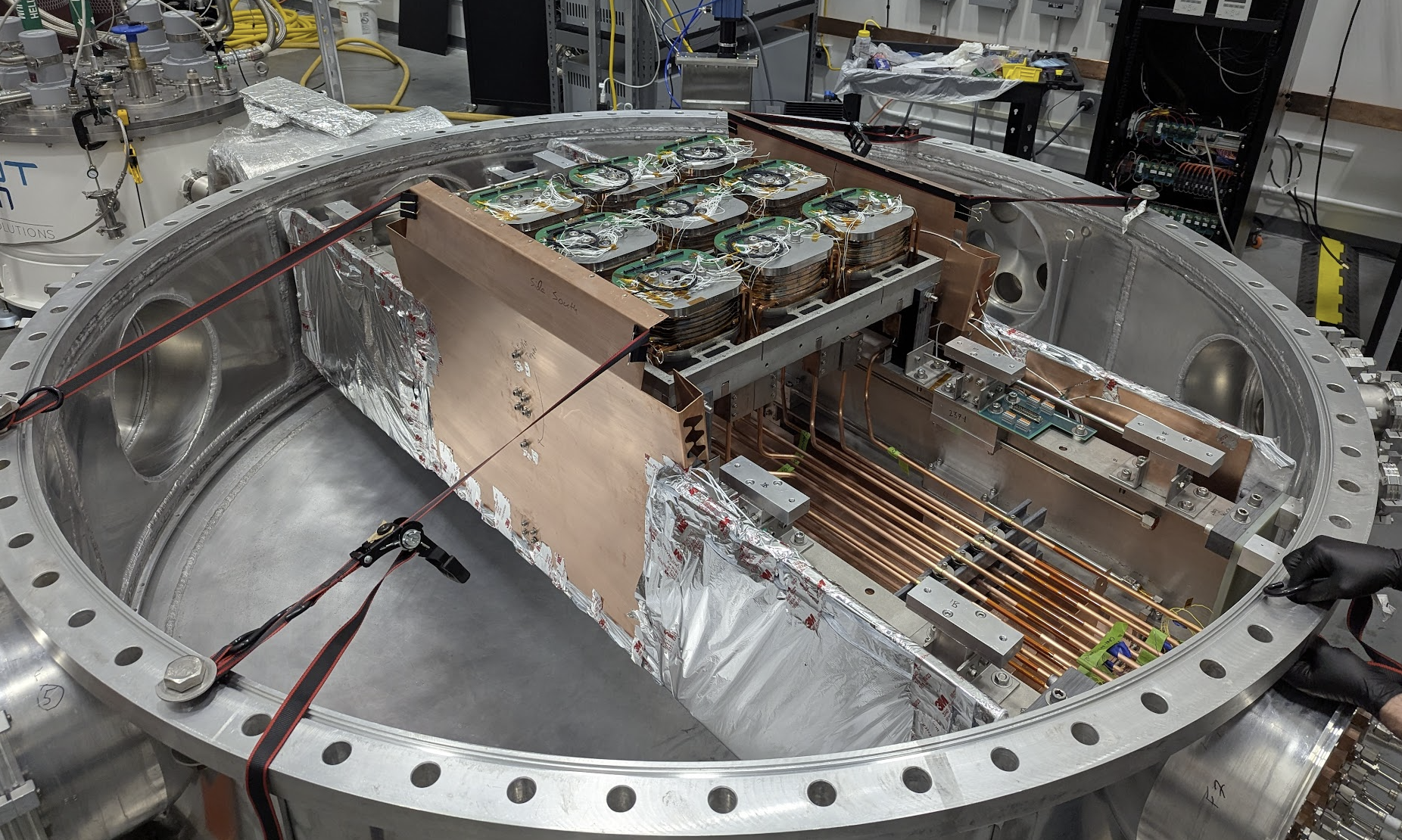}
  \caption{Photo of completed Canis 3x3 magnet array prior to cryostat closeout.}
  \label{fig_array_photo}
  \end{figure}

Following the completion of WP acceptance testing, the nine WPs of the Canis magnet array were installed onto the structural array plate and integrated into the Canis cryostat. The remaining cryostat and ex-vessel systems were integrated and the complete test cell was commissioned at 20 K. A photo of the Canis magnet array prior to cryostat closeout is shown in Fig. \ref{fig_array_photo}.

\subsection{Coil Characterization and Array Calibration}
The Canis magnet array test campaign required multiple layers of calibration to enable closed-loop field shaping control to a target of \(\leq\)1\% root mean square (RMS) field error. Aspects of the test campaign requiring calibration included the magnetic field measurement systems, electrical characterization of individual WPs at 20 K, and estimation of the ATLAS field scanning plane location and orientation with respect to the cold magnet array.

Two independent, pre-calibrated sensor systems were used to capture magnetic field with high accuracy. Nine Lake Shore HGCA-3020 cryogenic Hall effect sensors, with room temperature calibrations up to \(\pm\)30 kG were used to measure and control magnetic field in-vessel, local to each WP. Per the manufacturer, an approximate mean sensitivity shift of -0.4\% is expected for operation at 20 K, and this sensitivity scaling was applied globally to all HGCA-3020 sensors, regardless of individual calibration curve. For field measurement on the scanning plane, a Metrolab THM1176-MF 3-axis Hall probe, with 0.1 mT resolution and calibrated to \(\pm\)0.1\% up to 0.1 T was used, as described in section 5H.

\begin{figure}[!h]
  \centering
  \includegraphics[width=3.5in]{\figpath/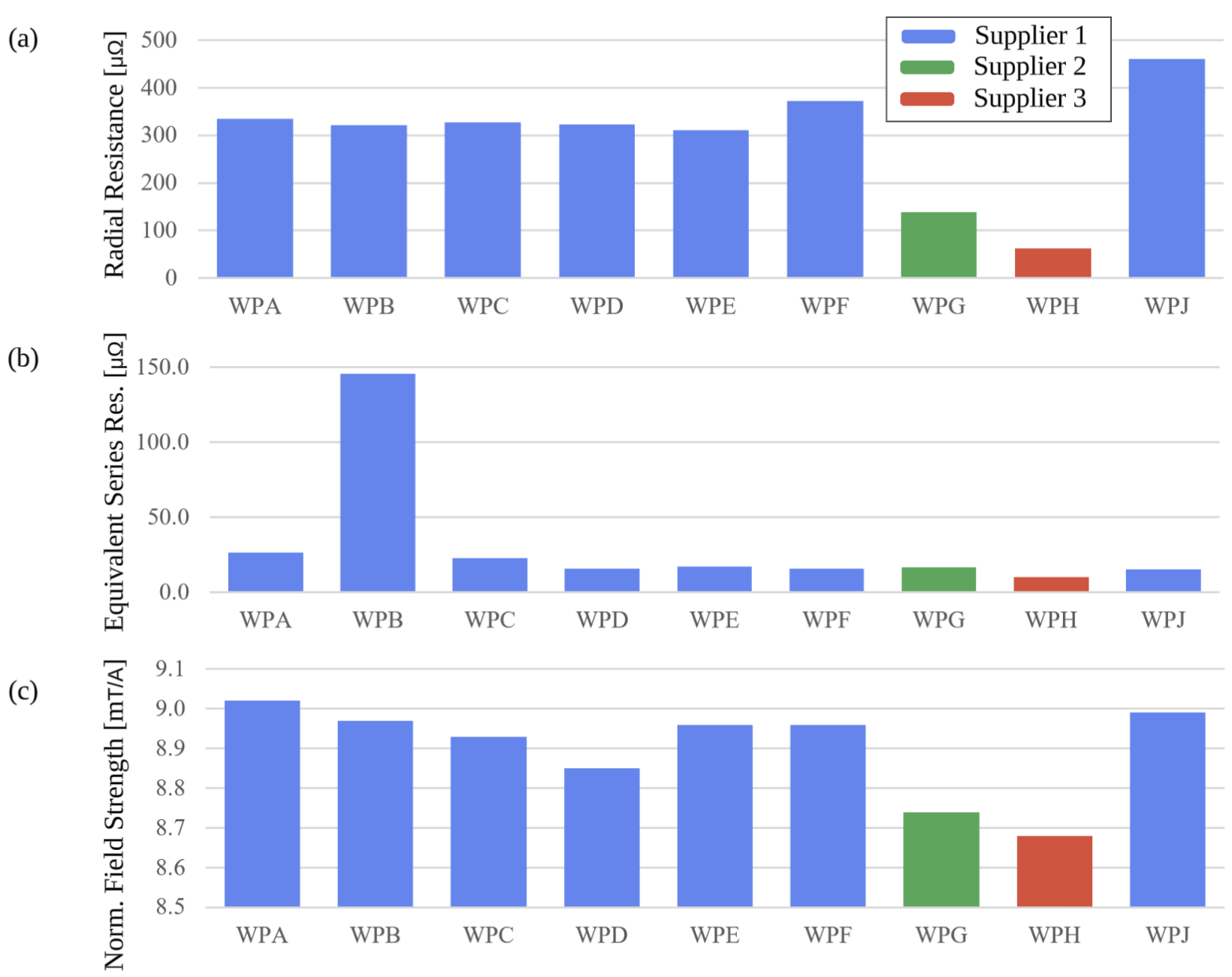}
  \caption{Winding pack (WP) performance trends over serial number as measured during 20 K testing, for WPs manufactured from three different suppliers. Shown are (a) radial resistance, (b) equivalent series resistance (ESR) and (c) field strength normalized by power supply current. Field is measured by a Lake Shore HGCA-3020 cryogenic Hall effect sensor, located along the central axis of the coil 36.3 mm from the coil midplane.}  \label{fig_wp_20k}
  \end{figure}

All nine WPs were independently electrically characterized at 20 K for radial resistance, ESR, and generated magnetic field normalized by that WP's power supply current. During this phase, only a single WP was charged and characterized at a time, with remaining WPs open circuited at the coupling box. The radial resistance of each WP was integrated into the control algorithm to enable closed-loop field control while limiting radial current heat power to a configurable value. 20 K WP performance is summarized in Fig. \ref{fig_wp_20k}.

During the array calibration phase, all current leads and other resistive components were characterized for resistance, heat power, and temperature rise. Additionally, the array calibration phase of the test program also enabled characterization of the 20 K helium cooling loop, including estimation of the helium circuit and cooling plate thermal efficiency, characteristic response times, and re-cool times.

\begin{figure}[!h]
  \centering
  \includegraphics[width=3.5in]{\figpath/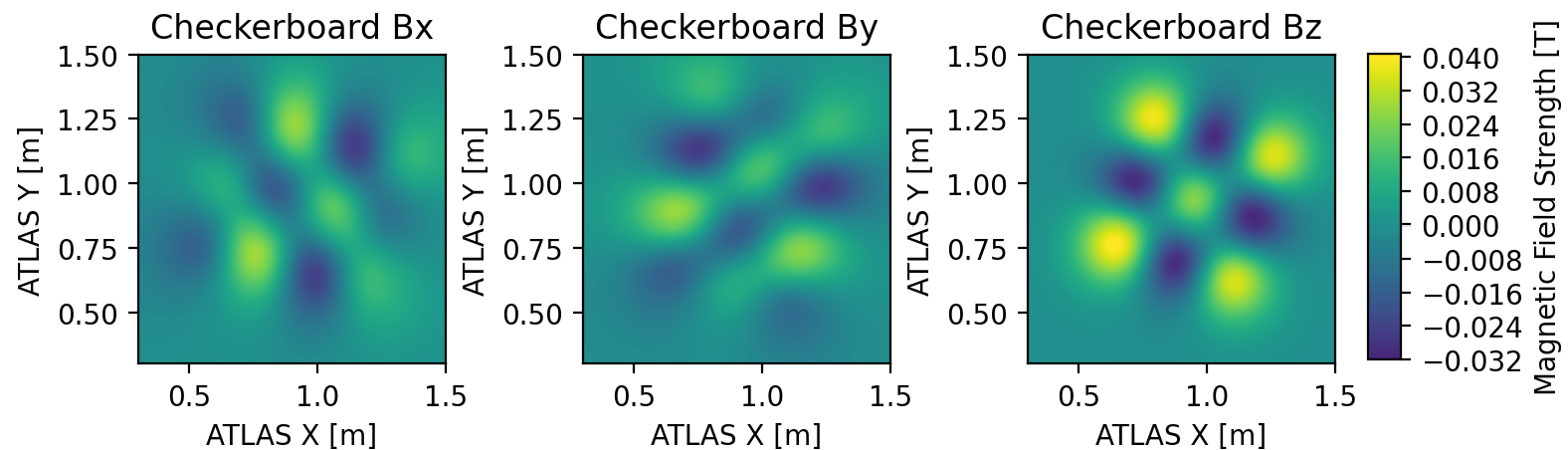}
  \caption{ATLAS scan of ``checkerboard'' pattern, showing $B_{x}$, $B_{y}$, and $B_{z}$ components of magnetic field measured on the scanning plane. Peak fields on the scanning plane are approximately \(\pm\)40 mT.}
  \label{fig_atlas_checkerboard}
  \end{figure}

Following individual coil characterization, the precise location and orientation of the ATLAS field scanning gantry with respect to the cold magnet array was estimated by scanning a reference field shape called the ``checkerboard'' field pattern. In this pattern, all magnets are charged to high-field in alternating polarity. A checkerboard pattern was chosen for scan plane calibration due to its large measurement range and field gradients relative to other field shapes. The checkerboard field used for calibrating the ATLAS scan plane was generated by controlling fields at each WP Hall sensor to WP-specific targets of approximately \(\pm\)1.25 T (corresponding to approximately \(\pm\)140 A in each coil), and the resultant field scan is shown in Fig. \ref{fig_atlas_checkerboard}. Note that field magnitudes on the 2-dimensional scan plane are in the \(\pm\)40 mT range and are lower than the \(\pm\)1.25 T range fields measured at each WP for closed-loop control. During this checkerboard field scan, the maximum field on HTS in the center coil was estimated to be 3.06 T.

\begin{figure}[!h]
  \centering
  \includegraphics[width=3.2in]{\figpath/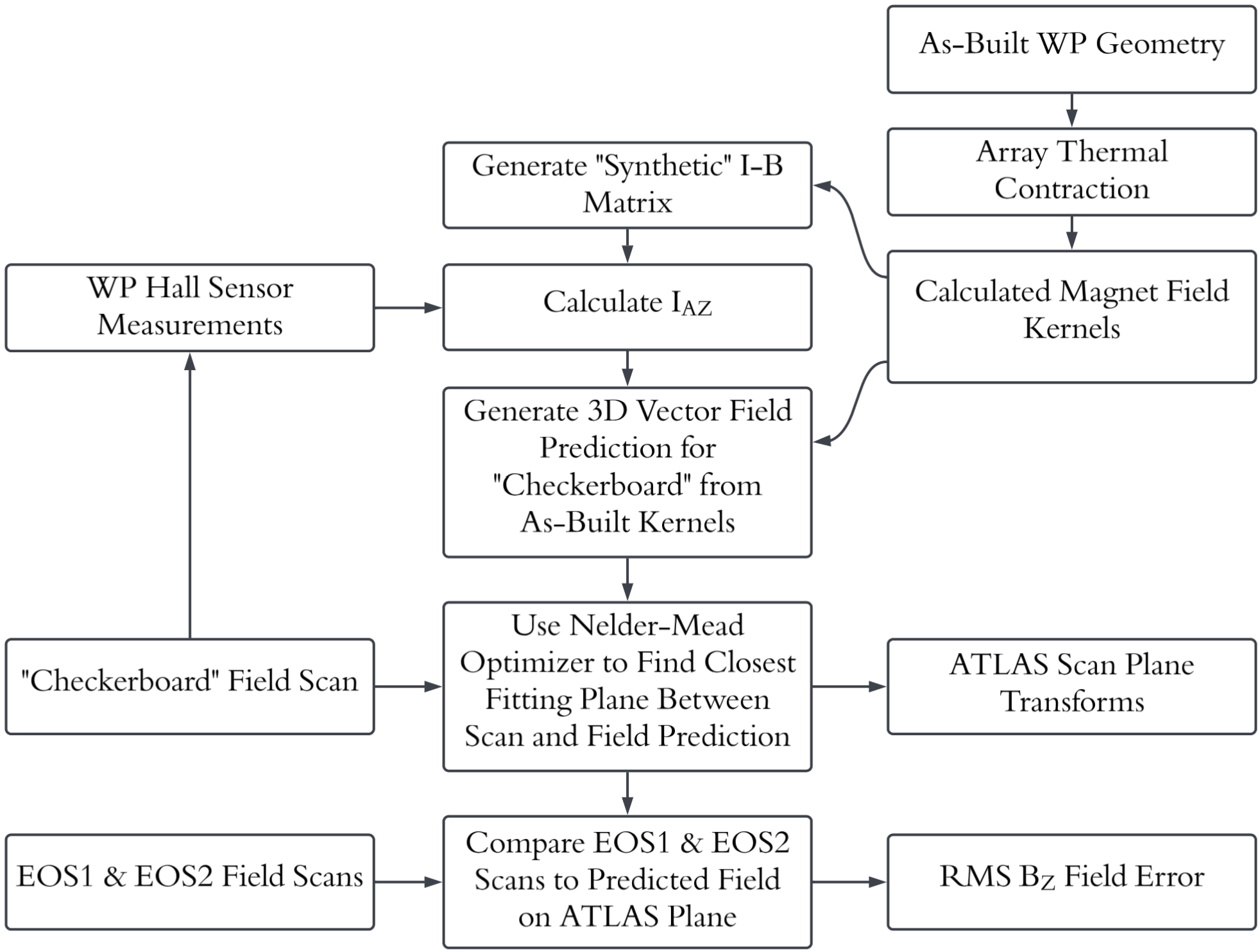}
  \caption{Summary of the workflow used to estimate ATLAS scan plane location.}
  \label{fig_atlas_workflow}
  \end{figure}

The workflow used to estimate the location and orientation of the ATLAS scanning plane based on the checkerboard calibration is summarized in Fig. \ref{fig_atlas_workflow}. The workflow began with generating 3-dimensional magnetic field kernels unique to each WP. These ``as-built'' field kernels considered the measured diameter of each constituent pancake, and are calculated using an efficient 1D integral method proposed by Landreman et al. \cite{landreman_field_2023}. Nine unique field kernels are calculated and superimposed, with the spacing between coils reduced from their nominal dimensions by the estimated thermal contraction of the magnet array plate at 20 K (\(\Delta L/L=-300.04\times 10^{-5}\))\cite{nist_316_properties}.

The aggregate 3-dimensional field created by superimposing nine unique field kernels was used to generate a ``synthetic'' I-B matrix $\mathbf{A}$, a 9x9 matrix predicting the measured magnetic field at each Hall sensor $\mathbf{B}$ due to the azimuthal current in each magnet $\mathbf{I_{az}}$. Using the inverted matrix $\mathbf{A}^{-1}$ and vector of field measurements $\mathbf{B}$, the azimuthal currents at each magnetic can be estimated per Eq. \ref{eq:IB_synth}:

\begin{equation}
  \begin{aligned}
  \mathbf{B} = \mathbf{A} \mathbf{I_{az}} \\
  \mathbf{I_{az}} = \mathbf{A}^{-1} \mathbf{B}
  \label{eq:IB_synth}
  \end{aligned}
  \end{equation}

The magnet array was operated in closed-loop control mode in the checkerboard pattern and the steady-state azimuthal currents in each magnet were estimated from Hall sensor measurements. These currents were used to generate a 3-dimensional field prediction within which the ATLAS scan plan could be located. An optimization routine based on the Nelder-Mead method was used to estimate nine degrees of freedom (DOF) associated with the ATLAS scan by minimizing vector RMS field error between actual scan measurements and a slice through the predicted field: 3 DOF for the XYZ position of the scan plane relative to the magnet array, 3 DOF for the angular rotation of the scan plane about the array XYZ axes, and 3 DOF for the angular orientation of the Metrolab THM1176-MF 3-axis Hall probe with respect to the scanning gantry axes. For the closed-loop field shaping campaign described in the following section, the 9 DOF of the ATLAS scanner were solved for and are shown in Table \ref{tab:atlas_loc}.

\begin{table}[h]
  \caption{\label{tab:atlas_loc} ATLAS orientation for field error estimation}
  \centering
  \begin{tabular}{l|c|c|c}\hline
  Group & \multicolumn{3}{c}{Degree of Freedom}\\
    & X & Y & Z \\\hline
  Scan Plane Offset [m] & -0.9508 & -0.9362 & 0.2418\\
  Scan Plane Rotation [\textdegree] & -0.0262 & 0.0254 & 72.70 \\
  Hall Probe Rotation [\textdegree] & -0.484 & -0.808 & 0.330 \\
  \end{tabular}
  \end{table}

From Table \ref{tab:atlas_loc} it can be seen that the ATLAS gantry was rotated 72.70{\textdegree} relative to the magnet array axes due to installation constraints. The scan plan was also estimated to be 8.2 mm closer to the magnet array than the nominal 25 cm.

Subsequent estimation of closed-loop field shaping error relied on both the calculated magnetic field kernels to predict 3-dimensional field from the array, and the optimized ATLAS scanning plane location and orientation to determine the plane onto which the field prediction should be projected for error calculation.

\subsection{Field Shaping Campaign}
The primary goal of the closed-loop field shaping phase of the Canis test program was to successfully generate multiple field shapes relevant to a planar coil stellarator with a field error $E_{RMS}$ as described by Eq. \ref{eq:bz_error} of less than 1\% RMS.

\begin{equation}
E_{RMS} = \frac{\sqrt{\frac{1}{N} \sum_{i=1}^{N} (B_{z,meas,i} - B_{z,pred,i})^2}}{|B_{z,pred,max}|}
\label{eq:bz_error}
\end{equation}

where $B_{z,meas,i}$ and $B_{z,pred,i}$ are the measured and predicted $B_{z}$ field at each ATLAS scan point, respectively, and $B_{z,pred,max}$ is the maximum predicted $B_{z}$ field on the ATLAS scan plane.

\begin{figure}[!h]
  \centering
  \includegraphics[width=3.5in]{\figpath/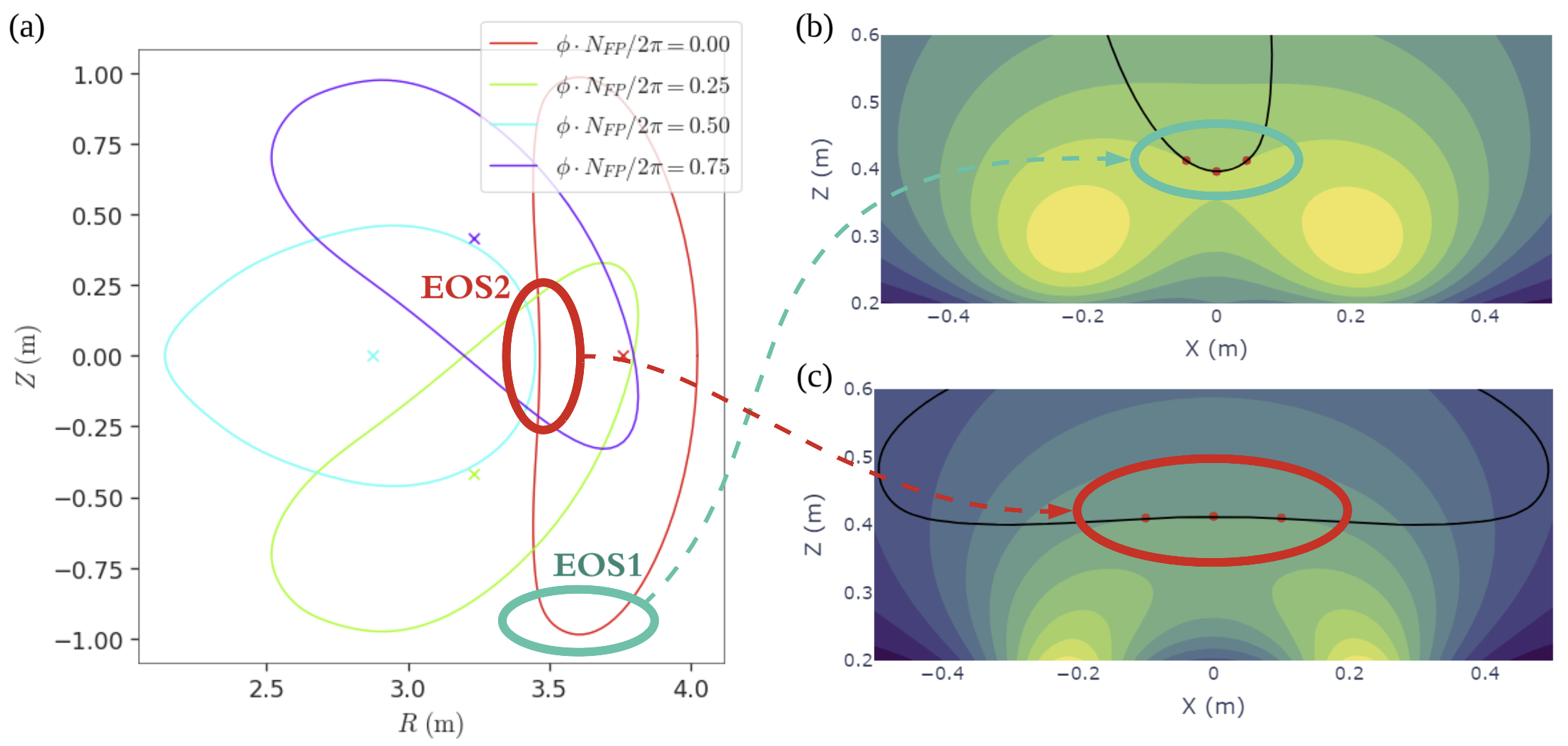}
  \caption{Workflow for generating Eos patches starts with identifying regions of curvature from Eos plasma equilibrium. (a) shows a plasma cross-sections including the $\phi=0$ section in red. (b) EOS1 and (c) EOS2 patches are generated by selectively matching the curvature of the $|B_{norm}|$ ($|B_{z}|$ for the Canis 3x3 magnet array) iso-surface at nine anchor points for the bottom and inboard segments of the $\phi=0$ section, respectively.}
  \label{fig_eos_patches}
  \end{figure}

Two specified field shapes, known as the ``EOS1'' and ``EOS2'' patches, were created by shaping the $|B_{z}|$ iso-surface of the Canis magnet array to match the last closed flux surface of a candidate Eos equilibrium. Specifically, the $|B_{z}|$ iso-surface of the Canis magnet array was fit to the location of nine anchor points on the $|B_{norm}|=0$ surface of Eos. The correspondence of the Canis magnet array $|B_{z}|$ iso-surface to the Eos $|B_{norm}|=0$ surface was chosen because a $|B_{norm}|=0$ surface from the Canis magnet array would be trivial in the absence of external fields. The two locations of interest that form the EOS1 and EOS2 patches correspond to the bottom and inboard regions of the plasma at the toroidal angle $\phi=0$, respectively. A cross-section of the Eos plasma indicating EOS1 and EOS2 patches is shown in Fig. \ref{fig_eos_patches}.

\begin{figure}[!h]
  \centering
  \includegraphics[width=3.2in]{\figpath/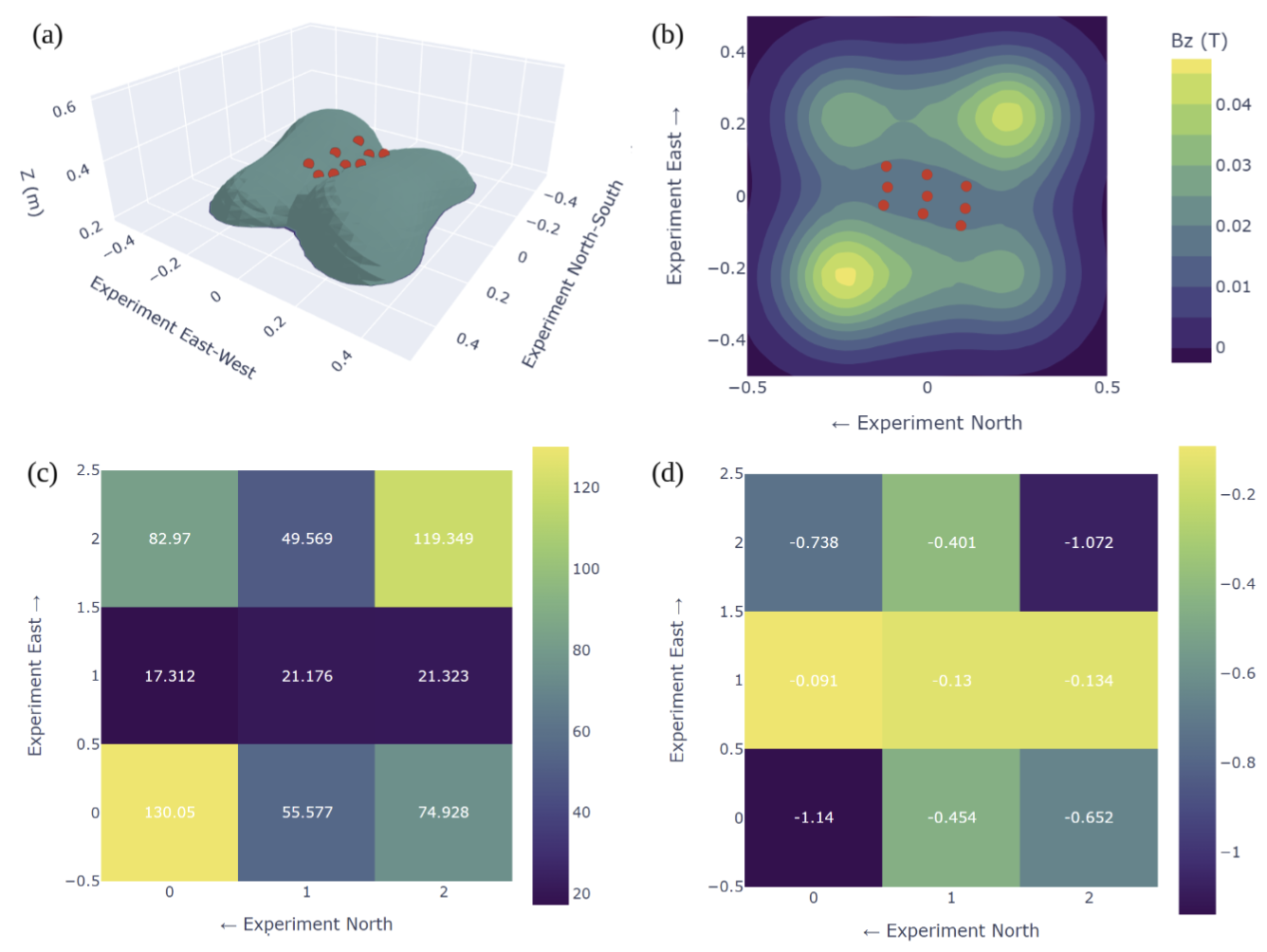}
  \caption{Detail for EOS1 patch, including (a) isometric view of $|B_{z}|$ iso-surface curvature showing nine anchor points, (b) estimated $|B_{z}|$ as projected onto the scanning plane, (c) expected azimuthal current in each WP, and (d) predicted field as measured at each WP's Hall effect sensor.}
  \label{fig_eos1}
  \end{figure}

\begin{figure}[!h]
  \centering
  \includegraphics[width=3.2in]{\figpath/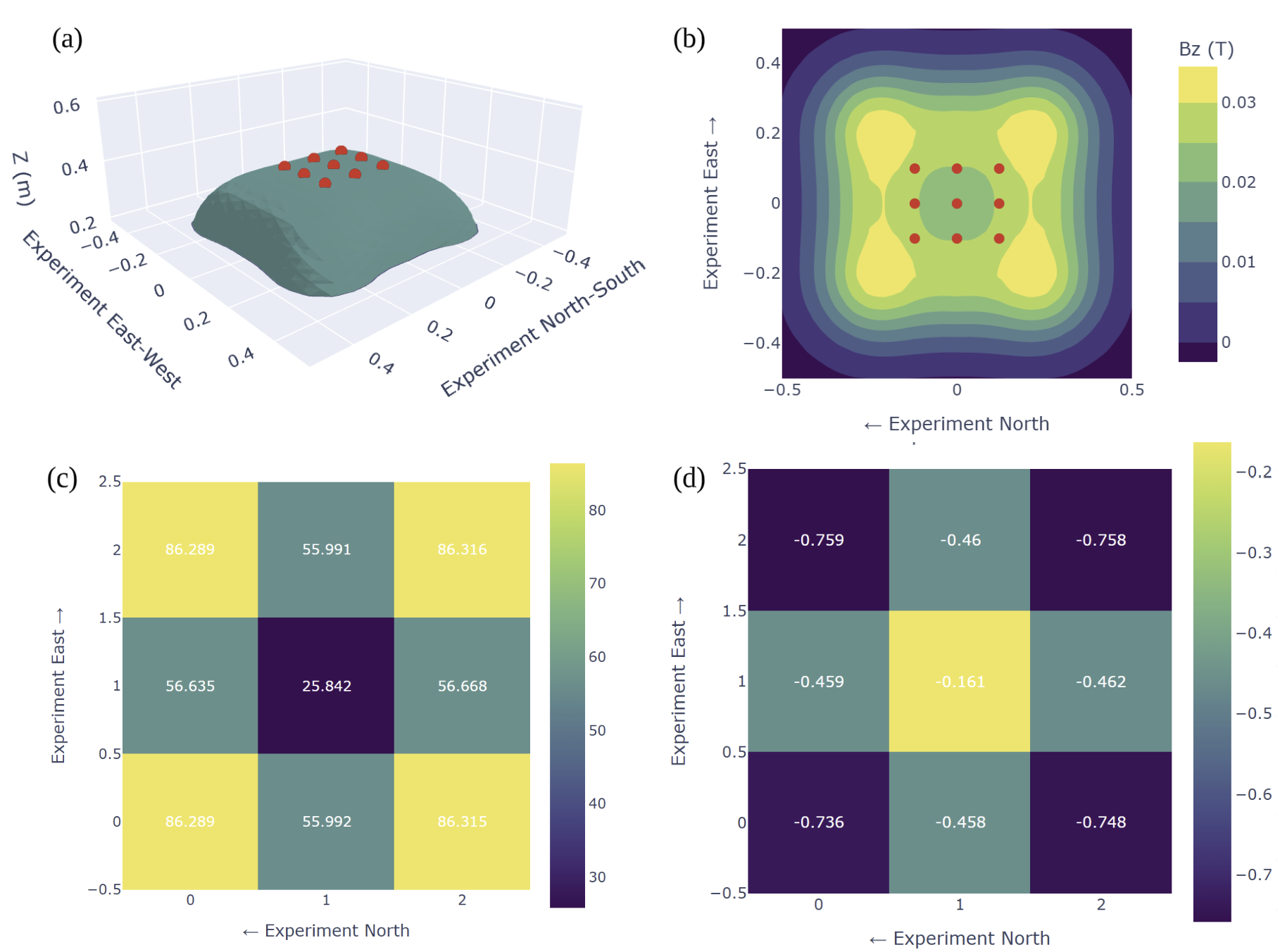}
  \caption{Detail for EOS2 patch, including (a) isometric view of $|B_{z}|$ iso-surface curvature showing nine anchor points, (b) estimated $|B_{z}|$ as projected onto the scanning plane, (c) expected azimuthal current in each WP, and (d) predicted field as measured at each WP's Hall effect sensor.}
  \label{fig_eos2}
  \end{figure}

For both EOS1 and EOS2 field shapes, a solution for the azimuthal current in each of the nine WPs was calculated from the intended magnetic field at the iso-surface anchor points. From this solution, the expected magnetic field as projected onto a plane 25 cm above the array midplane can be calculated, the expected Hall sensor measurement at each WP. An isometric view of the $|B_{z}|$ iso-surfaces and anchor points, field projections onto the scanning plane, ideal WP azimuthal currents, and expected Hall effect field measurements at each WP are shown for the EOS1 and EOS2 patches in Fig. \ref{fig_eos1} and Fig. \ref{fig_eos2}, respectively.

\begin{figure}[!h]
  \centering
  \includegraphics[width=3.0in]{\figpath/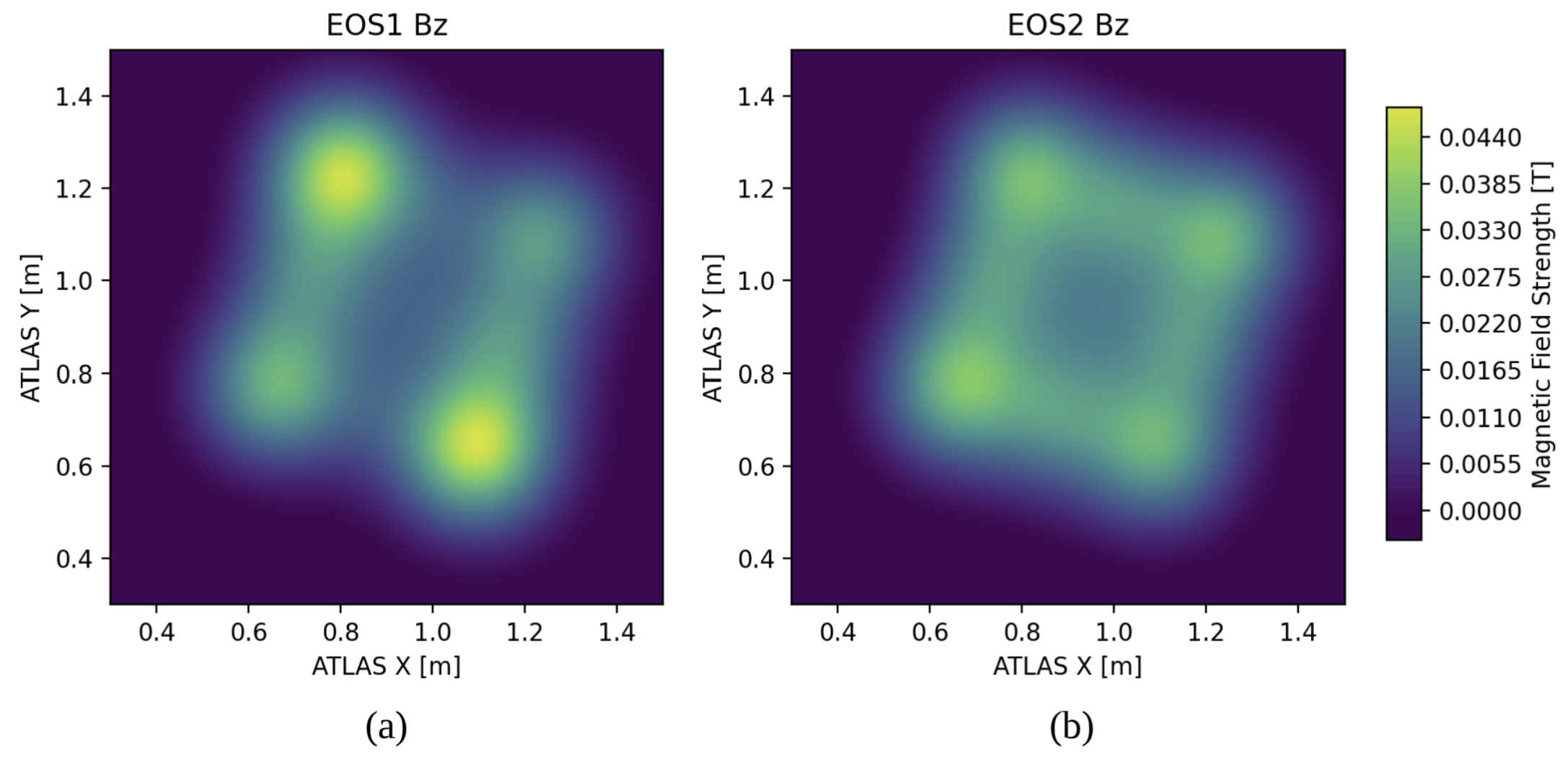}
  \caption{$B_{z}$ field measurement from ATLAS scans for (a) EOS1 and (b) EOS2 field shapes. The maximum $B_{z}$ field measured in the EOS1 field shape was approximately 47.2 mT.}
  \label{fig_eos_scans}
  \end{figure}

The Canis magnet array was operated in closed-loop field control mode to generate the EOS1 and EOS2 patches. In this mode, the control system adjusts the commanded power supply current for each coil to minimize error between the field measured at the Hall sensor in each WP and the field targets from Fig. \ref{fig_eos1}(d) and Fig. \ref{fig_eos2}(d) for EOS1 and EOS2, respectively. When the array had reached steady state, the magnetic field on the scanning plane was measured with the ATLAS field scanner with 12.5 mm resolution in both scanning axes, as is shown for both EOS1 and EOS2 in Fig. \ref{fig_eos_scans}. Based on the translational and rotational transform of the scanning plane with respect to the magnet array from Table \ref{tab:atlas_loc}, new predictions for magnetic field on the actual scanning plane were generated.

\begin{figure}[!h]
  \centering
  \includegraphics[width=3.5in]{\figpath/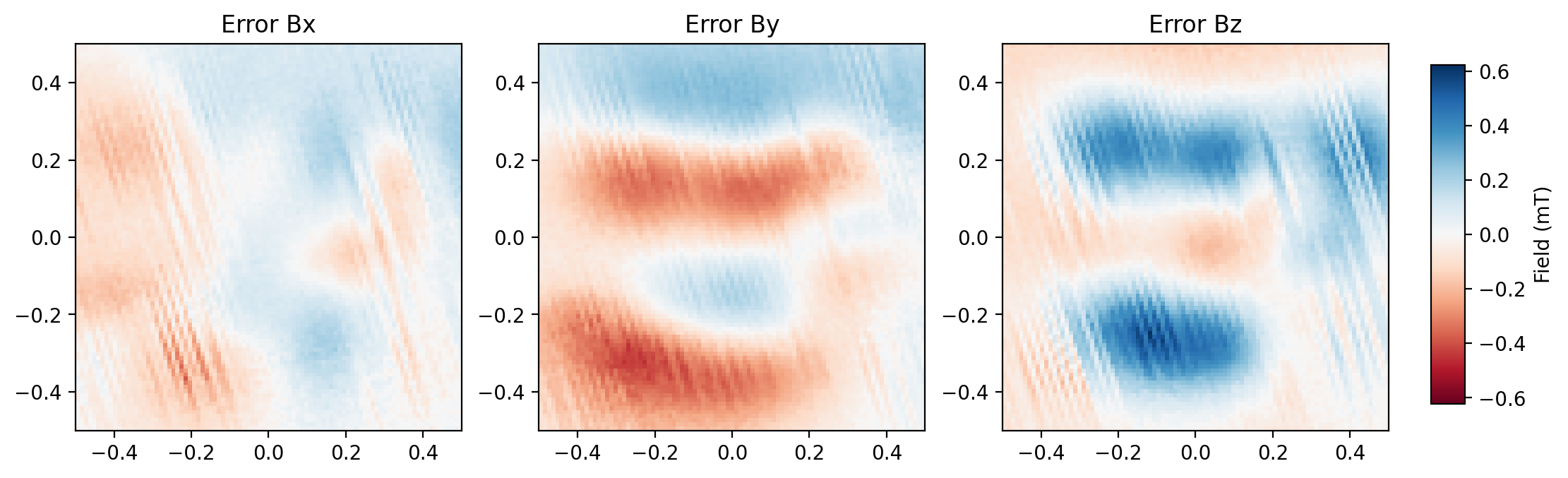}
  \caption{$B_{x}$, $B_{y}$, and $B_{z}$ field error for EOS1 field shape. Error is computed between the field measured by ATLAS and the predicted field projected onto the ATLAS plane as decribed in Fig. \ref{fig_atlas_workflow}.}
  \label{fig_eos1_error}
  \end{figure}

\begin{figure}[!h]
  \centering
  \includegraphics[width=3.5in]{\figpath/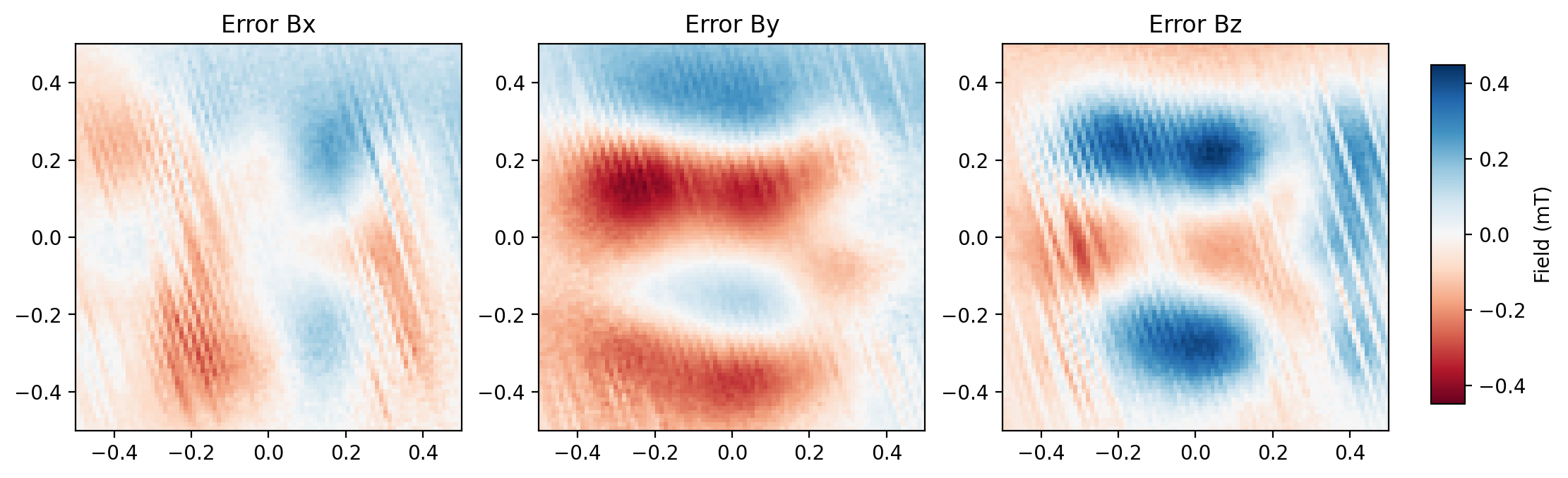}
  \caption{$B_{x}$, $B_{y}$, and $B_{z}$ field error for EOS2 field shape. Error is computed between the field measured by ATLAS and the predicted field projected onto the ATLAS plane as decribed in Fig. \ref{fig_atlas_workflow}.}
  \label{fig_eos2_error}
  \end{figure}

The difference in $B_{x}$, $B_{y}$, and $B_{z}$ components of magnetic field between the field prediction and the ATLAS scan are shown for EOS1 and EOS2 patches in Fig. \ref{fig_eos1_error} and Fig. \ref{fig_eos2_error}, respectively. Field error $E_{RMS}$ for each field shape was calculated per Eq. \ref{eq:bz_error} by evaluating the RMS error between prediction and measurement of $B_{z}$, normalized by the maximum predicted field on the scan plane. RMS error, peak $\mathbf{B}$ vector magnitude error, and peak $B_{z}$ error for both field shapes are summarized in Table \ref{tab:error_results}.

\begin{table}[h]
  \caption{\label{tab:error_results} Field error estimation for EOS1 and EOS2 scans}
  \centering
  \begin{tabular}{l|c|c}\hline
  Error & \multicolumn{2}{c}{Field Shape}\\
    & EOS1 & EOS2 \\\hline
  $E_{RMS}$ [\%] (mT) & 0.56\% (0.27 mT) &  0.60\% (0.24 mT) \\
  Peak $|\mathbf{B}|$ Error [\%] (mT) & 1.38\% (0.66 mT) & 1.18\% (0.48 mT) \\
  Peak $B_{z}$ Error [\%] (mT) & 1.33\% (0.62 mT) & 1.17\% (0.45 mT)\\
  \end{tabular}
  \end{table}

A characteristic diagonal streaking can be seen in the error plots in Figs. \ref{fig_eos1_error} and \ref{fig_eos2_error}. These streaks are aligned with the x-axis of the ATLAS scanning gantry and correspond to lines of constant y-position. Integrating error between position feedback from the ATLAS linear encoders and from the stepper drive motor encoders over the full y-axis of travel showed no net deviation between those measurements, suggesting positional error was limited to specific raster lines and was not accruing globally. This phenomenon was likely attributed to slight error in motor position feedback control at particular y-positions while scanning along the x-axis.

\begin{figure}[!h]
  \centering
  \includegraphics[width=3.5in]{\figpath/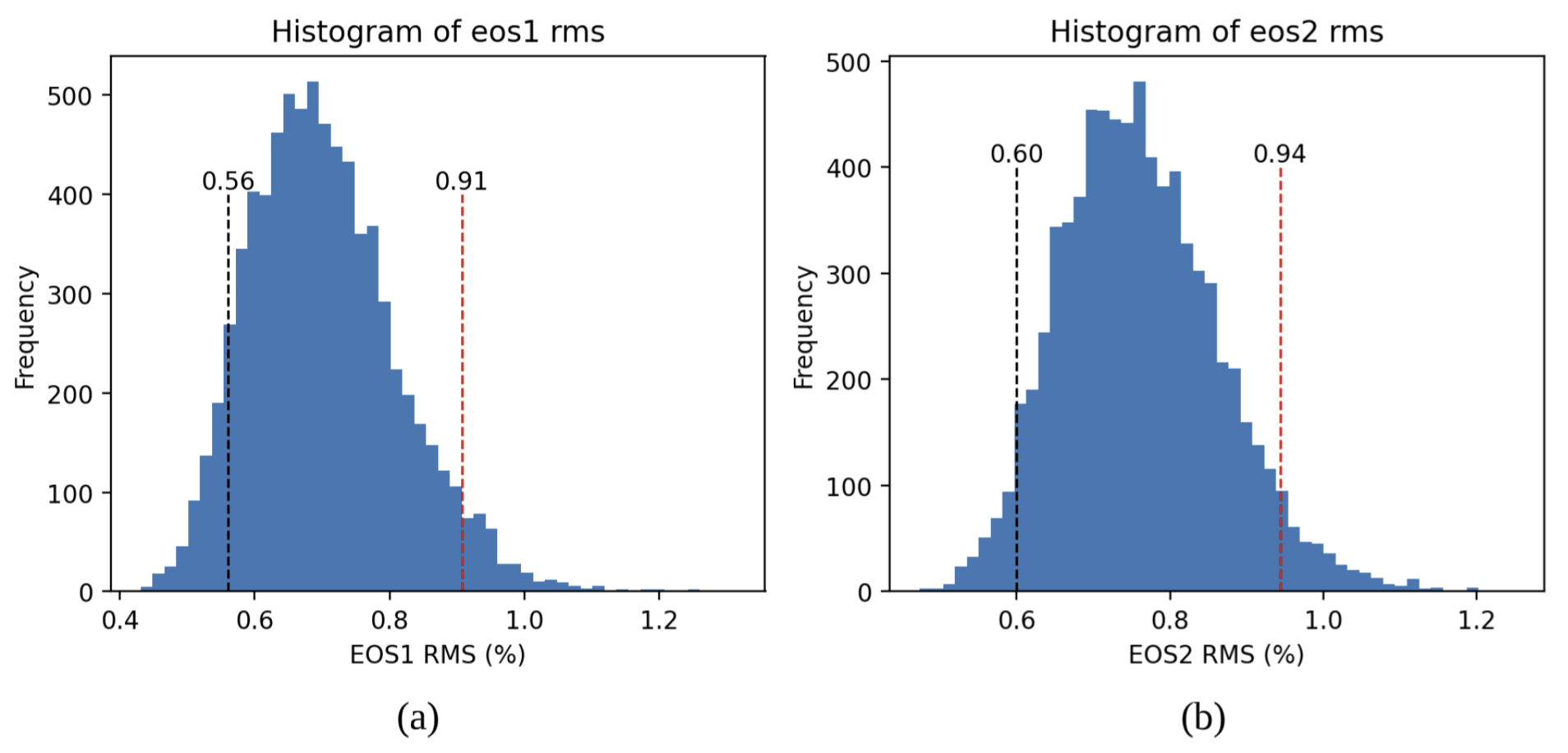}
  \caption{Histogram of resulting $E_{RMS}$ values for (a) EOS1 and (b) EOS2 when WP Hall sensor measurements and WP x-y locations are perturbed in a normal distribution based on their known uncertainties. Each Monte Carlo analysis executed 7,578 simulation runs. The nominal $E_{RMS}$ for EOS1 and EOS2 as shown in Table \ref{tab:error_results} are indicated by the dashed black lines. The dashed red lines indicate the 95\textsuperscript{th} percentile results for the EOS1 and EOS2 simulations, which are 0.91\% and 0.94\%, respectively.}
  \label{fig_error_hist}
  \end{figure}

Potentially significant sources of error in the generation of field predictions and ATLAS scan plane orientations included end-to-end uncertainty in the WP Hall sensor measurements (\(\pm\)0.9\%) and variability in the installed and cold position of each magnet, estimated at \(\pm\)0.5 mm in either array x-axis or y-axis. The \(\pm\)0.9\% uncertainty in Hall sensor measurements includes uncertainty in the calibration and data acquisition chain, and uncertainty in the precise z-coordinate location and angular orientation of the installed Hall sensor. A Monte Carlo analysis in which WP Hall measurements and magnet positions were perturbed in a normal distribution and propagated through the full analysis workflow in Fig. \ref{fig_atlas_workflow} was run to estimate uncertainty in the $E_{RMS}$ values in Table \ref{tab:error_results}. The $E_{RMS}$ results from 7,578 perturbed simulations are summarized in the histograms in Fig. \ref{fig_error_hist}. For EOS1 and EOS2, the Monte Carlo simulations yielded a 95\textsuperscript{th} percentile value for $E_{RMS}$ of 0.91\% and 0.94\%, respectively.

\section{Discussion}
The Canis magnet array demonstrated the control of magnetic field projected onto a plane within \(\pm\)1\% of predictions. In an integrated planar coil stellarator, the shaping coil array is composed of hundreds of planar shaping coils which are arranged onto a non-planar surface that is roughly conformal to the plasma shape. Although the Canis magnet array includes only nine magnets in a 3x3 grid in a fully planar arrangement (i.e. mounted onto a flat surface), the array provides several key features demonstrating its relevance to the planar coil array, and provides a fundamental proof of concept for field shaping via arrays of planar coils.

First, the mutual inductance between coils is shown analytically to fall off to 1 mH and below (\(\leq\)1\% of self inductance) for two magnets separated by a magnet between them. From this, we can consider the center magnet to be maximally coupled only to the adjacent coils, and mutual coupling beyond the adjacent coils to be negligible. Demonstrating closed-loop field control with a single magnet maximally coupled to its neighbors in a stellarator-relevant geometry suggests that this approach is viable for larger arrays of coils.

Second, the planar nature of the Canis magnet array as compared to a conformal shaping coil array for a planar coil stellarator has minimal effect on the inductive coupling between magnets, particularly for magnets separated by a magnet between them. Further, generating highly non-uniform fields of high curvature is more challenging with a planar array, because the arrangement of the coils does not naturally provide a curvature that conforms to the field shape. A planar 3x3 arrangement of coils was selected to appropriately capture inductive coupling between magnets for at least one magnet while minimizing the physical size and cost of the magnet array. 

Third, although the Canis magnet array doesn't precisely recreate optimized stellarator field shapes, forces, or currents as would be seen in an Eos-scale shaping coil array, the non-uniform fields generated by the Canis magnet array have a similar qualitative character to Eos field shapes, show similar strong gradients of current and magnetic field, and require similar field shaping accuracy to implement.

Limitations of the Canis magnet array's applicability to a planar coil stellarator include that the array cannot produce a net toroidal flux, and therefore cannot reproduce the field-line following, closed flux surfaces, or quasisymmetry that will be seen in the Eos integrated stellarator. Developing a system capable of generating the toroidal flux seen in the Eos shaping coil array would have significantly increased the cost, schedule, and complexity of the Canis program. For a single prototypical FSU or the Eos stellarator, much more dense integration of structural, cooling, and power systems will need to be developed than were scoped and implemented for the Canis program.

\section{Summary and Future Work}
Thea Energy successfully prototyped and tested the Canis 3x3 magnet array, demonstrating the array's suitability for generating 3-dimensional field shapes representative of what is required for a planar coil stellarator. Nine planar shaping coils were manufactured, individually tested, and integrated into a purpose-built cryogenic test infrastructure at Thea Energy. Measurements of stellarator-relevant field shapes generated by the magnet array indicated that the magnetic field projected onto a plane 25 cm from the array was controlled to within \(\pm\)1\% of predictions.

Additional planned testing of the Canis 3x3 magnet array includes demonstration of transient field shape control, confirmation of the self protecting behavior of the HTS shaping coils during various quench scenarios, and  robustness to variability and defects in the manufacturing and integration of planar shaping coils.

\section*{Acknowledgments}
Thea Energy would like to thank Andrey Lednev and Derek Fluegge for their contributions to the control and safety system infrastructure. We would also like to thank Jeffrey Whalen and Tom Painter of MagCorp\textsuperscript{\texttrademark{}} for their consultation regarding winding techniques and pancake joint design.

This research was funded by Thea Energy and performed as part of the DOE Milestone Based Fusion Development Program (DE-SC0024881).


\bibliographystyle{IEEEtran}
\bibliography{planar_coil_magnet_array}

\begin{thebibliography}{10}
\providecommand{\url}[1]{#1}
\csname url@samestyle\endcsname
\providecommand{\newblock}{\relax}
\providecommand{\bibinfo}[2]{#2}
\providecommand{\BIBentrySTDinterwordspacing}{\spaceskip=0pt\relax}
\providecommand{\BIBentryALTinterwordstretchfactor}{4}
\providecommand{\BIBentryALTinterwordspacing}{\spaceskip=\fontdimen2\font plus
\BIBentryALTinterwordstretchfactor\fontdimen3\font minus
  \fontdimen4\font\relax}
\providecommand{\BIBforeignlanguage}[2]{{%
\expandafter\ifx\csname l@#1\endcsname\relax
\typeout{** WARNING: IEEEtran.bst: No hyphenation pattern has been}%
\typeout{** loaded for the language `#1'. Using the pattern for}%
\typeout{** the default language instead.}%
\else
\language=\csname l@#1\endcsname
\fi
#2}}
\providecommand{\BIBdecl}{\relax}
\BIBdecl

\bibitem{gates_stellarator_2025}
\BIBentryALTinterwordspacing
D.~Gates, S.~Aslam, B.~Berzin, P.~Bonofiglo, A.~Cote, D.~Dudt, E.~Flom,
  D.~Fort, A.~Koen, T.~Kruger, S.~Kumar, M.~Martin, A.~Ottaviano, S.~Pasmann,
  P.~Romano, C.~Swanson, L.~Tang, E.~Winkler, and R.~Wu,
  ``\BIBforeignlanguage{en}{Stellarator fusion systems enabled by arrays of
  planar coils},'' \emph{\BIBforeignlanguage{en}{Nuclear Fusion}}, vol.~65,
  no.~2, p. 026052, Jan. 2025, publisher: IOP Publishing. [Online]. Available:
  \url{https://dx.doi.org/10.1088/1741-4326/ada56c}
\BIBentrySTDinterwordspacing

\bibitem{swanson_scoping_2025}
\BIBentryALTinterwordspacing
C.~Swanson, D.~Gates, S.~Kumar, M.~Martin, T.~Kruger, D.~Dudt, P.~Bonofiglo,
  and t.~T.~E. team, ``\BIBforeignlanguage{en}{The scoping, design, and plasma
  physics optimization of the {Eos} neutron source stellarator},''
  \emph{\BIBforeignlanguage{en}{Nuclear Fusion}}, vol.~65, no.~2, p. 026053,
  Jan. 2025, publisher: IOP Publishing. [Online]. Available:
  \url{https://dx.doi.org/10.1088/1741-4326/ada56a}
\BIBentrySTDinterwordspacing

\bibitem{planar_coil_stellarator_patent}
``Planar coil stellarator,'' Patent PCT/US2023/063\,949, March 8, 2023,
  published as WO/2023/178004.

\bibitem{bosch_lessons_2011}
H.-S. Bosch, ``Lessons learned in the construction of {Wendelstein} 7-{X},'' in
  \emph{Proceedings of {SOFE} 2011}, Chicago, IL, Jun. 2011.

\bibitem{bosch_engineering_2018}
H.-S. Bosch, T.~Andreeva, R.~Brakel, T.~Bräuer, D.~Hartmann, A.~Holtz,
  T.~Klinger, H.~Laqua, M.~Nagel, D.~Naujoks, K.~Risse, A.~Spring, T.~S.
  Pedersen, T.~Rummel, P.~van Eeten, A.~Werner, and R.~Wolf, ``Engineering
  {Challenges} in {W7}-{X}: {Lessons} {Learned} and {Status} for the {Second}
  {Operation} {Phase},'' \emph{IEEE Transactions on Plasma Science}, vol.~46,
  no.~5, pp. 1131--1140, May 2018, conference Name: IEEE Transactions on Plasma
  Science.

\bibitem{neilson_lessons_2009}
G.~Neilson, C.~Gruber, J.~Harris, D.~Rej, R.~Simmons, and R.~Strykowsky,
  ``Lessons learned in risk management on {NCSX},'' in \emph{2009 23rd
  {IEEE}/{NPSS} {Symposium} on {Fusion} {Engineering}}, Jun. 2009, pp. 1--6,
  iSSN: 2155-9953.

\bibitem{chrzanowski_lessons_2009}
\BIBentryALTinterwordspacing
J.~H. Chrzanowski, T.~G. Meighan, S.~Raftopolous, L.~Dudek, and P.~J. Fogarty,
  ``\BIBforeignlanguage{English}{Lessons {Learned} {During} the {Manufacture}
  of the {NCSX} {Modular} {Coils}},'' Princeton Plasma Physics Lab. (PPPL),
  Princeton, NJ (United States), Tech. Rep. PPPL-4442, Sep. 2009. [Online].
  Available: \url{https://www.osti.gov/biblio/963972}
\BIBentrySTDinterwordspacing

\bibitem{almagri_hsx_1999}
\BIBentryALTinterwordspacing
A.~Almagri, D.~Anderson, F.~Anderson, P.~Probert, J.~Shohet, and J.~Talmadge,
  ``A helically symmetric stellarator ({HSX}),'' \emph{IEEE Transactions on
  Plasma Science}, vol.~27, no.~1, pp. 114--115, Feb. 1999. [Online].
  Available: \url{http://ieeexplore.ieee.org/document/763074/}
\BIBentrySTDinterwordspacing

\bibitem{kruger_coil_2025}
\BIBentryALTinterwordspacing
T.~Kruger, M.~Martin, D.~Gates, and t.~T.~E. Team,
  ``\BIBforeignlanguage{en}{Coil optimization methods for a planar coil
  stellarator},'' \emph{\BIBforeignlanguage{en}{Nuclear Fusion}}, vol.~65,
  no.~2, p. 026051, Jan. 2025, publisher: IOP Publishing. [Online]. Available:
  \url{https://dx.doi.org/10.1088/1741-4326/ada56b}
\BIBentrySTDinterwordspacing

\bibitem{molodyk_development_2021}
\BIBentryALTinterwordspacing
A.~Molodyk, S.~Samoilenkov, A.~Markelov, P.~Degtyarenko, S.~Lee, V.~Petrykin,
  M.~Gaifullin, A.~Mankevich, A.~Vavilov, B.~Sorbom, J.~Cheng, S.~Garberg,
  L.~Kesler, Z.~Hartwig, S.~Gavrilkin, A.~Tsvetkov, T.~Okada, S.~Awaji,
  D.~Abraimov, A.~Francis, G.~Bradford, D.~Larbalestier, C.~Senatore,
  M.~Bonura, A.~E. Pantoja, S.~C. Wimbush, N.~M. Strickland, and A.~Vasiliev,
  ``\BIBforeignlanguage{en}{Development and large volume production of
  extremely high current density {YBa2Cu3O7} superconducting wires for
  fusion},'' \emph{\BIBforeignlanguage{en}{Scientific Reports}}, vol.~11,
  no.~1, p. 2084, Jan. 2021, number: 1 Publisher: Nature Publishing Group.
  [Online]. Available: \url{https://www.nature.com/articles/s41598-021-81559-z}
\BIBentrySTDinterwordspacing

\bibitem{hartwig_sparc_2023}
\BIBentryALTinterwordspacing
Z.~S. Hartwig, R.~F. Vieira, D.~Dunn, T.~Golfinopoulos, B.~LaBombard, C.~J.
  Lammi, and P.~C. Michael, ``The {SPARC} {Toroidal} {Field} {Model} {Coil}
  {Program},'' \emph{IEEE Transactions on Applied Superconductivity}, pp.
  1--18, 2023, conference Name: IEEE Transactions on Applied Superconductivity.
  [Online]. Available: \url{https://ieeexplore.ieee.org/document/10316582}
\BIBentrySTDinterwordspacing

\bibitem{sanabria_development_2024}
\BIBentryALTinterwordspacing
C.~Sanabria, A.~Radovinsky, C.~L. Craighill, K.~Uppalapati, and A.~Warner,
  ``Development of a high current density, high temperature superconducting
  cable for pulsed magnets,'' \emph{Superconductor Science and Technology},
  vol.~37, no.~11, Oct. 2024. [Online]. Available:
  \url{https://iopscience.iop.org/article/10.1088/1361-6668/ad7efc}
\BIBentrySTDinterwordspacing

\bibitem{schwartz_quench_2014}
\BIBentryALTinterwordspacing
J.~Schwartz, ``Quench in high temperature superconductor magnets,'' 2014,
  arXiv:1401.3937 [physics]. [Online]. Available:
  \url{http://arxiv.org/abs/1401.3937}
\BIBentrySTDinterwordspacing

\bibitem{laan_delamination_2007}
\BIBentryALTinterwordspacing
D.~C. v.~d. Laan, J.~W. Ekin, C.~C. Clickner, and T.~C. Stauffer,
  ``\BIBforeignlanguage{en}{Delamination strength of {YBCO} coated conductors
  under transverse tensile stress*},''
  \emph{\BIBforeignlanguage{en}{Superconductor Science and Technology}},
  vol.~20, no.~8, p. 765, Jun. 2007. [Online]. Available:
  \url{https://dx.doi.org/10.1088/0953-2048/20/8/007}
\BIBentrySTDinterwordspacing

\bibitem{lu_rebco_2025}
\BIBentryALTinterwordspacing
J.~Lu, J.~Levitan, A.~Hutley, and H.~Bai, ``{REBCO} {Delamination}
  {Characterization} by 90 {Degree} {Peel} {Test},'' \emph{IEEE Transactions on
  Applied Superconductivity}, vol.~35, no.~5, pp. 1--5, Aug. 2025, conference
  Name: IEEE Transactions on Applied Superconductivity. [Online]. Available:
  \url{https://ieeexplore.ieee.org/document/10767303?utm_source=chatgpt.com}
\BIBentrySTDinterwordspacing

\bibitem{ballarino_1999}
\BIBentryALTinterwordspacing
A.~Ballarino, ``High {Temperature} {Superconducting} {Current} {Leads} for the
  {Large} {Hadron} {Collider},'' \emph{IEEE Transactions on Appiled
  Superconductivity}, vol.~9, no.~2, Jun. 1999. [Online]. Available:
  \url{https://cds.cern.ch/record/409769/files/cer-000336910.pdf}
\BIBentrySTDinterwordspacing

\bibitem{hahn_hts_2011}
S.~Hahn, D.~K. Park, J.~Bascunan, and Y.~Iwasa, ``{HTS} {Pancake} {Coils}
  {Without} {Turn}-to-{Turn} {Insulation},'' \emph{IEEE Transactions on Applied
  Superconductivity}, vol.~21, no.~3, pp. 1592--1595, Jun. 2011, conference
  Name: IEEE Transactions on Applied Superconductivity.

\bibitem{lu_contact_2018}
\BIBentryALTinterwordspacing
J.~Lu, J.~Levitan, D.~McRae, and R.~Walsh, ``\BIBforeignlanguage{en}{Contact
  resistance between two {REBCO} tapes: the effects of cyclic loading and
  surface coating},'' \emph{\BIBforeignlanguage{en}{Superconductor Science and
  Technology}}, vol.~31, no.~8, p. 085006, Jul. 2018, publisher: IOP
  Publishing. [Online]. Available:
  \url{https://dx.doi.org/10.1088/1361-6668/aacd2d}
\BIBentrySTDinterwordspacing

\bibitem{hahn_no-insulation_2013}
\BIBentryALTinterwordspacing
S.~Hahn, Y.~Kim, D.~Keun~Park, K.~Kim, J.~P. Voccio, J.~Bascuñán, and
  Y.~Iwasa, ``\BIBforeignlanguage{en}{No-insulation multi-width winding
  technique for high temperature superconducting magnet},''
  \emph{\BIBforeignlanguage{en}{Applied Physics Letters}}, vol. 103, no.~17, p.
  173511, Oct. 2013. [Online]. Available:
  \url{https://pubs.aip.org/apl/article/103/17/173511/26425/No-insulation-multi-width-winding-technique-for}
\BIBentrySTDinterwordspacing

\bibitem{lecrevisse_metal-as-insulation_2018}
\BIBentryALTinterwordspacing
T.~Lécrevisse, A.~Badel, T.~Benkel, X.~Chaud, P.~Fazilleau, and P.~Tixador,
  ``Metal-as-insulation variant of no-insulation {HTS} winding technique:
  pancake tests under high background magnetic field and high current at 4.2
  {K},'' \emph{Superconductor Science and Technology}, vol.~31, no.~5, p.
  055008, May 2018. [Online]. Available:
  \url{https://iopscience.iop.org/article/10.1088/1361-6668/aab4ec}
\BIBentrySTDinterwordspacing

\bibitem{song_over-current_2015}
\BIBentryALTinterwordspacing
J.-B. Song, S.~Hahn, T.~Lécrevisse, J.~Voccio, J.~Bascuñán, and Y.~Iwasa,
  ``Over-current quench test and self-protecting behavior of a 7 {T}/78 mm
  multi-width no-insulation {REBCO} magnet at 4.2 {K},'' \emph{Superconductor
  Science and Technology}, vol.~28, no.~11, p. 114001, Nov. 2015. [Online].
  Available:
  \url{https://iopscience.iop.org/article/10.1088/0953-2048/28/11/114001}
\BIBentrySTDinterwordspacing

\bibitem{choi_characteristic_2016}
\BIBentryALTinterwordspacing
J.~Choi, S.~K. Kim, S.~Kim, K.~Sim, M.~Park, and I.~K. Yu, ``Characteristic
  {Analysis} of a {Sample} {HTS} {Magnet} for {Design} of a 300 {kW} {HTS} {DC}
  {Induction} {Furnace},'' \emph{IEEE Transactions on Applied
  Superconductivity}, vol.~26, no.~3, pp. 1--5, Apr. 2016. [Online]. Available:
  \url{http://ieeexplore.ieee.org/document/7400948/}
\BIBentrySTDinterwordspacing

\bibitem{seungyong_hahn_no-insulation_2012}
\BIBentryALTinterwordspacing
{Seungyong Hahn}, {Dong Keun Park}, J.~Voccio, J.~Bascunan, and Y.~Iwasa,
  ``No-{Insulation} ({NI}) {HTS} {Inserts} for \${\textgreater}\$1 {GHz}
  {LTS}/{HTS} {NMR} {Magnets},'' \emph{IEEE Transactions on Applied
  Superconductivity}, vol.~22, no.~3, pp. 4\,302\,405--4\,302\,405, Jun. 2012.
  [Online]. Available: \url{http://ieeexplore.ieee.org/document/6099575/}
\BIBentrySTDinterwordspacing

\bibitem{lecrevisse_metal-as-insulation_2022}
\BIBentryALTinterwordspacing
T.~Lécrevisse, X.~Chaud, P.~Fazilleau, C.~Genot, and J.-B. Song,
  ``\BIBforeignlanguage{en}{Metal-as-insulation {HTS} coils},''
  \emph{\BIBforeignlanguage{en}{Superconductor Science and Technology}},
  vol.~35, no.~7, p. 074004, May 2022, publisher: IOP Publishing. [Online].
  Available: \url{https://dx.doi.org/10.1088/1361-6668/ac49a5}
\BIBentrySTDinterwordspacing

\bibitem{mun_electrical_2020}
J.~Mun, C.~Lee, K.~Sim, C.~Lee, M.~Park, and S.~Kim, ``Electrical
  {Characteristics} of {Soldered} {Metal} {Insulation} {REBCO} {Coil},''
  \emph{IEEE Transactions on Applied Superconductivity}, vol.~30, no.~4, pp.
  1--4, Jun. 2020, conference Name: IEEE Transactions on Applied
  Superconductivity.

\bibitem{tang_zethus_2025}
K.~Tang, ``Zethus: a prototype {HTS} planar shaping coil for the {Eos}
  stellarator,'' in \emph{Magnet Technology 29}, Jul. 2025, abstract submitted
  January 15, 2025.

\bibitem{erickson_multiparticle_1970}
K.~Erickson, ``Multiparticle production in liquid hydrogen and carbon by
  charged cosmic ray hadrons of energy greater than 70 {GeV},'' The University
  of Michigan, Technical {Report} UM HE 70-4, Apr. 1970.

\bibitem{landreman_field_2023}
\BIBentryALTinterwordspacing
M.~Landreman, S.~Hurwitz, and T.~M. Antonsen~Jr, ``Efficient calculation of
  self magnetic field, self-force, and self-inductance for electromagnetic
  coils. {II}. {Rectangular} cross-section,'' Oct. 2023, arXiv:2310.12087
  [physics]. [Online]. Available: \url{http://arxiv.org/abs/2310.12087}
\BIBentrySTDinterwordspacing

\bibitem{nist_316_properties}
\BIBentryALTinterwordspacing
N.~I. of~Standards and Technology, ``Material properties: 316 stainless,''
  2024. [Online]. Available:
  \url{https://trc.nist.gov/cryogenics/materials/316Stainless/316Stainless_rev.htm}
\BIBentrySTDinterwordspacing

\end{thebibliography}

\newpage

\vspace{11pt}

\vfill

\end{document}